# Advanced Deuterium Fusion Rocket Propulsion For Manned Deep Space Missions

Friedwardt Winterberg, Department of Physics, Mail Stop # 0220, LP 204
University of Nevada, Reno, Phone# (775) 784-6789, Fax# (775) 784-1398
Email : winterbe@physics.unr.edu

**Abstract**

Excluding speculations about future breakthrough discoveries in physics, it is shown that with what is at present known, and also what is technically feasible, manned space flight to the limits of the solar system and beyond deep into the Oort cloud is quite well possible. Using deuterium as the rocket fuel of choice, abundantly available on the comets of the Oort cloud, rockets driven by deuterium fusion, can there be refueled. To obtain a high thrust with a high specific impulse, favors the propulsion by deuterium micro-bombs, and it is shown that the ignition of deuterium micro-bombs is possible by intense GeV proton beams, generated in space by using the entire spacecraft as a magnetically insulated billion volt capacitor. The cost to develop this kind of propulsion system in space would be very high, but it can also be developed on earth by a magnetically insulated Super Marx Generator. Since the ignition of deuterium is theoretically possible with the Super Marx Generator, rather than deuterium-tritium with a laser where 80% of the energy goes into neutrons, would also mean a breakthrough in fusion research, and therefore would justify the large development costs.

## Preface

My interest in space flight can be traced back to the time I was about 10 years old, when as a birthday gift I got a popular book about the feasibility of space flight. There I heard for the first time about Oberth and Goddard, and of the possibility to reach the moon with a multistage rocket. It was the same time when Hahn and Strassmann had announced the discovery of nuclear fission with the possibility of an atomic bomb by a fission chain reaction.

Having been born in Germany in 1929, I received my PhD. in physics under Heisenberg in 1955. Inspired by the 15 Megaton hydrogen bomb test, conducted in 1952 by the United States, I have been since 1954 deeply interested in the non–fission ignition of thermonuclear reactions by inertial confinement. At this time all fusion research in US was still classified, but I had quite independently discovered the basic principles of inertial confinement, the Guderley convergent shock wave, and the Rayleigh imploding shell solutions. In 1956 I presented my findings in Goettingen, at a meeting at the Max Planck Institute, which was organized by Von Weizsäcker. The abstracts of this meeting still exist and are kept in the library of University of Stuttgart.

Because in 1958, I had delivered at the 2nd United Nations Conference on the Peaceful Use of Atomic Energy, a paper which turned out to be the importance for Nerva-type nuclear rocket reactors, I was under "Operation Paperclip" invited by the US government to come to the United States. In San Diego I met Ted Taylor and Freeman Dyson, who were working on the famous "Orion" nuclear bomb propulsion concept. This concept is generally credited to Ulam, but as I know from conversations I had with Heisenberg, a likewise idea was presented to Heisenberg by Wernher von Braun, who had visited Heisenberg in Berlin in or around 1942. Because of my idea to use the Guderly convergent shock wave solution for thermonuclear ignition, Ted Taylor and Freeman Dyson were interested in my joining their group. But because at that time this work was classified and I was not yet a US citizen, this was not possible.

About 10 years later, in 1967, I saw a new possibility for the non-fission ignition of thermonuclear micro-explosions by intense relativistic electron and ion beams, driven by a high voltage Marx Generator. This ignition concept could be used not only for the controlled release of energy by nuclear fusion, but also for the propulsion of a space craft, replacing the pusher plate of the Orion concept with a magnetic mirror, reflecting the plasma-fireball of the thermonuclear micro-explosion [1, 2].

This idea was adopted by the British Interplanetary Society in their 1978 Project Daedalus starship study, replacing neutron-rich deuterium-tritium (DT) thermonuclear explosive with a neutron-poor $DHe^3$ explosive [3]. Unlike the DT reaction where 80% of the released energy goes into neutrons, most of the energy in $D\text{-}He^3$ reaction goes into alpha particles which can be deflected by a magnetic mirror. But since $He^3$ is not everywhere abundantly available, it was proposed to “mine” it from the atmosphere of Jupiter.

Studies to propel a space-craft with the matter-antimatter annihilation reaction have also been made, but because of the enormous technical problem to produce antimatter in appreciable quantities, this can be delegated into the realm of science fiction, as are space-warp drives, space flight through wormholes and other fantasies not supported by one shred of experimental evidence. Only the idea to use nano-gram amounts of antimatter for the ignition of fission-fusion micro-explosions appears to have some credible potential, but even there the production and storage of nano-gram quantities of antimatter poses serious technical problems [4].

We have no reason to expect that new fundamental laws in physics, which could lead to a breakthrough in propulsion, are still awaiting us. Very much as America was discovered only once, it is quite well possible that all the fundamental laws of physics relevant for propulsion have been discovered, challenging our imagination to find out if they are sufficient to invent propulsion systems which ultimately might bring us to earthlike planets of nearby solar systems.

I will end this preface with an imaginary talk by Ted Taylor to Freeman Dyson as it has been recorded by George Dyson, the son of Freeman Dyson, in his book “Project Orion – The True Story of the Atomic Spaceship” [5], followed by a dream of Ted Taylor. “Freeman’s hope for the Orion had rested on the fact that there seems to be no law of nature forbidding the construction of fission-free bombs”, and on the belief that “improvements in the design of the nuclear devices (by reducing the fraction of total yield due to fission) might achieve reduction factors of $10^2$ to $10^3$. This belief in small, fission-free bombs has largely evaporated”. “One exception is Ted (Taylor). He remains convinced that small, clean bombs could propel Orion, -but he still fears, more than ever that such devices would be irresistible as weapons, until we outgrow the habit of war. There are lots of different routes to that final result of a very clean bomb”, he says: “could you make a one-kiloton explosion in which the fission yield was zero, which is bad news on the proliferation front, but could turn Orion into something quite clean?” “Freeman thinks Ted is wrong – and Ted hopes Freeman is right”. I for my part think Freeman is wrong.

Many years later shortly before his death, Ted reports: “I had a dream last night, about a new form of nuclear weapon, and I am really scared of it”. He tells us when he woke up, he wrote down his dream, and it appeared scientifically sound and feasible. What was it? We never will know with certainty but I have a guess. It is the possibility of chemical super explosives, (explained in Appendix) powerful enough to ignite a thermonuclear bomb.

## 1. Introduction

As Hermann Oberth did prove in 1923, in his book "The Rocket into Planetary Space" [6] for chemical rockets, I will try to prove for thermonuclear rockets the following:

1. At the present state of science and technology one can build spaceships driven by deuterium thermonuclear reactions, able to reach the outer limits of the solar system.
2. Such spaceships permit the manned exploration of the entire solar system and beyond, with the ultimate potential to reach nearby solar systems.
3. The cost in research and development to build such spaceships will be high, but still well within what is economically feasible.
4. Using the same physical principles as for deuterium fusion rockets, will also lead to the realization of clean nuclear energy, justifying the expenditures for these large projects.

With deuterium as the rocket fuel of choice, and any inert material suitable as a propellant, both available on most planetary bodies, but in particular on the comets of the Oort cloud, the idea is that by a gradual radial expansion from the sun, by building bridges over the Oort clouds presumably surrounding all suns, earthlike planets in neighboring solar systems can eventually be reached.

Towards this goal the first and most important step is to reach the focus of the Einstein gravitational lens at 550 AU. At this location one can use the sun as the lens of a super telescope, an idea first proposed by Claudio Maccone in 1993. We now know that there are planets in nearby solar systems, and very likely earthlike planets. It is that only with this gigantic telescope we are able to determine if life on these planets is possible. But because of the complexity for this task, a manned mission to the Einstein gravitational lens focus is likely to be needed, possible only with advanced nuclear rocket propulsion.

## 2. On deuterium, argon ion lasers and keV super explosives

The goal is a space craft which can be refueled while landing on a planetary body, which can be a planet, an asteroid or a comet. With heavy water in water abundantly available on many planetary bodies, but in particular on comets, suggests to use deuterium as the thermonuclear rocket fuel. The ignition of deuterium though is more difficult than the ignition of the deuterium-tritium (DT) reaction, or the deuterium-helium3 ($DHe^3$) reaction.

The DT reaction is the easiest to ignite, but there 80% of the energy goes into neutrons, which can not be deflected by a magnetic mirror. In the $DHe^3$ reaction all the energy goes into charged fusion products, but in a mixture of D with $He^3$ there are still some neutron producing DD reactions. More important is the fact that unlike deuterium, $He^3$ is largely unavailable. There is some indication for $He^3$ on the surface of moon. In the Daedalus starship study of the British Interplanetary Society, it was proposed to "mine" $He^3$ from the atmosphere of Jupiter. In either case the cost to recover appreciable amounts of $He^3$ would be very high.

For the DD reaction the situation is quite different. Because there the instantaneous burn with deuterium of the T- and $He^3$- reaction products of deuterium, makes possible a detonation wave in dense deuterium. In this detonation wave only 38% of the energy released goes into neutrons, unlike the 80% for the DT reaction.

Deuterium can be extracted from water with relative ease in three steps:

1. Water is electrolytically split into hydrogen and oxygen.
2. The hydrogen gas composed of $H_2$ and HD is cooled down until it liquifies, whereby the heavier HD is separated by the force of gravity from the lighter $H_2$.
3. The newly produced HD is heated up and passed through a catalyst, splitting HD into $H_2$ and $D_2$, according to the equation [7]:

$$2HD \longrightarrow H_2 + D_2$$

Since the gravitational field on the surface of a comet or small planet, from where the $D_2$ shall be extracted, is small, the apparatus separating the liquid HD from $H_2$ must be set into rapid rotation.

The comparatively small amount of energy needed for the separation can ideally be drawn from a ferroelectric capacitor (for example a barium-titanate capacitor with a dielectric constant $\varepsilon \approx 5000$), to be charged up to many kilovolts by a small fraction of the electric energy drawn from the deuterium fusion explosions through a magneto hydrodynamic loop [2]. One can also draw this energy from a small on-board nuclear reactor requiring only a small radiator, slowly charging the capacitor. Alternatively, one may store the needed energy in the magnetic field of a superconductor.

For the launching of the spacecraft into earth orbit a very different scheme is proposed. It requires special materials are readily available on earth, but not on extraterrestrial bodies serving as landing points to refuel the spacecraft. There primarily water as a source of deuterium is needed.

In the Orion bomb propulsion project a large number of fission bombs, or fission-triggered fusion bombs, were proposed to lift the spacecraft into space. Since this would release a large amount of highly radioactive fission products into the atmosphere, it was one of the causes which killed Orion. Even though large payloads can be brought into earth orbit by chemical rockets, this remains very expensive and an acceptable less expensive nuclear alternative is highly desirable.

There seem to be two possibilities:

1. A laser driven by a high explosive, powerful enough to ignite a DT micro-explosion, which in turn can launch a thermonuclear detonation in deuterium [8].
2. The second possibility is more speculative: It is the conjectured existence of chemical keV superexplosives. These are chemical compounds formed under high pressure, resulting in keV bridges between inner electron shells, able to release intense bursts of keV X-rays, capable of igniting a DT thermonuclear reaction, which in turn could by propagating burn ignite a larger deuterium detonation.

For the realization of the first possibility, one may consider pumping a solid argon rod with a convergent cylindrical shock wave driven by a high explosive [9]. If the argon rod is placed in the center of convergence to reach a temperature of 90,000 ˚ K, this will populate in the argon the upper ultraviolet laser level, remaining frozen in the argon during its following rapid radial expansion. The energy thusly stored in the upper laser level can then be removed by a small Q-switched laser from the rod in one run into a powerful laser pulse, to be optically focused onto a thermonuclear target.

For the realization of the second possibility, one would have to subject suitable materials to very high pressure [10,11]. These energetic states can only be reached if during their compression the materials are not appreciably heated, because such heating would prevent the electrons from forming the bridges between the inner electron shells. Details of the second possibility will be given in the appendix.

## 3. On magnetic insulation and inductive charging

There are two concepts which are of great importance for the envisioned realization of a deuterium fusion driven starship:

1. The concepts of magnetic insulation. It permits the attainment of ultrahigh voltages in high vacuum [1].
2. The concept of inductive charging, by which a magnetically insulated conductor can be charged up to very high electric potentials [12].

**To 1.**

In a greatly simplified way magnetic insulation can be understood as follows: If the electric field on the surface of a negatively charged conductor reaches a critical field of the order $E_c \sim 10^7$ *V/cm*, the conductor becomes the source of electrons emitted by field emission. The critical electric field for the emission of ions from a positively charged conductor is ~ $10^8$ V/cm. Therefore, if in a high voltage diode the electric field reaches ~ $10^7$ V/cm, breakdown will occur by electric field emission from the cathode to the anode. But if a magnetic field of strength $B$ measured in Gauss, is applied in a direction parallel to the negatively charged surface, and if $B > E$, where $E$ (like $B$) is measured in electrostatic cgs units, the field emitted electrons make a drift motion parallel to the surface of the conductor with the velocity

$$v_d = c\frac{\underline{E} \times \underline{B}}{B^2} = c\frac{E}{B} \tag{1}$$

To keep $v_d/c < 1$ then requires that $E < H$. Let us assume that $H \approx 2 \times 10^4$ G, which can be reached with ordinary electromagnets, mean that $E \le 2 \times 10^4$ esu = $6 \times 10^6$ V/cm. For a conductor with a radius of $\ell \sim 10\text{m} = 10^3$ cm, an example for a small starship configuration, one can reach a voltage of the order $E\ell \le 6 \times 10^9$ Volts.

**To 2.**

To charge the spacecraft to the required gigavolt potentials we choose for its architecture a large, but hollow cylinder, which at the same time serves to act as a large magnetic field coil. If on the inside of this coils thermionic electron emitters are placed, and if the magnetic field of the coil rises in time, Maxwell's equation $curl\,\underline{E} = -(1/c)\partial\underline{B}/\partial t$ induces inside the coil an azimuthal electric field:

$$E_\phi = -\frac{r}{2c}\dot{B}_z \tag{2}$$

where $B = B_z$ is directed along the z-axis, with ν the radial distance from the axis of the coil. In combination with the axial magnetic field, the electrons from the thermionic emitters make a radial inward directed motion

$$v_r = c\frac{E_\phi}{B_z} = -\frac{r}{2}\frac{\dot{B}_z}{B_z} \tag{3}$$

By Maxwell's equation $div\,\underline{E} = 4\pi ne\underline{r}$, this leads to the buildup of an electron cloud inside the cylinder resulting in the radial electric field

$$E_r = 2\pi ner \tag{4}$$

This radial electric field leads to an additional azimuthal drift motion

$$v_\phi = c\,E_r/B_z \tag{5}$$

superimposed on the radially directed inward drift motion $v_r$.

For the newly formed electron cloud to be stable, its maximum electron number density must be below the Brillouin limit:

$$n < n_{max} = B^2/4\pi\, mc^2 \tag{6}$$

where $mc^2 = 8.2 \times 10^{-7}$ erg is the electron rest mass energy. For $B = 2\times10^4$G one finds that $n_{\max} \approx 4\times10^{13}$ cm$^{-3}$. To reach with a cylindrical electron cloud of radius $R$ a potential equal to $V$, requires an electron number density $n \approx V/\pi eR^2$. For $V = 10^9$ volts $\approx 3\times10^6$ esu and $R = 10^3$cm, one finds that $n \sim 2\times10^9$ cm$^{-3}$, well below $n_{max}$.

## 4. On deuterium as the preferred nuclear rocket fuel

To appreciate the importance of deuterium as the preferred and abundantly available nuclear rocket fuel, one must consider the secondary reactions of the $He^3$ and T D-D fusion reaction products with D. Taking these reactions into account one obtains from 6 deuterium nuclei an energy of 26.8 MeV in charged fusion products, made up from $He^3$ and H, and an energy of 16.55 MeV in neutrons. This means that 62% of the energy is released into charged fusion products and 38% into neutrons. This is a substantial improvement over the DT reaction where only 20% of the energy goes into $He^4$.

Of interest is also the average velocity, averaged over the momentum of the charged fusion products, because it is a measure of the maximum specific impulse, respectively the maximum exhaust velocity:

$$\bar{v} = \frac{\sum_{i=1}^{6} m_i v_i}{\sum_{i=1}^{6} m_i} \tag{7}$$

For the six charged fusion reaction products (given in **table 1**) one obtains $\bar{v} = 1.5 \times 10^9$ cm/s.

| Fusion Product | Energy [MeV] | Velocity $10^9$ [cm/s] |
|---|---|---|
| $He^3$ | 0.8 | 1.23 |
| H | 3.0 | 2.40 |
| H | 14.7 | 5.30 |
| $He^4$ | 3.6 | 1.31 |
| $He^4$ | 3.7 | 1.33 |
| T | 1.0 | 0.80 |

**Table 1:** The charged fusion products of a detonation in deuterium**:** Their energy and velocity

Because the 6 charged fusion products are accompanied by 9 electrons, they have to share their kinetic energy with 9 electrons. This reduces the maximum specific impulse by the factor $\sqrt{(6+9)/6} = \sqrt{2.5}$ to $\bar{v} = 0.95 \times 10^9$cm/s.

A reduction of the specific impulse does not occur if the electrons have enough time to escape the burning plasma behind the detonation front that is in a time shorter than the time for them to be heated up by the charged fusion products.

The time needed for the electrons to be heated by the charged fusion reaction products can be computed from the range $\lambda_0$ of the charged fusion products and their velocity $v$, in plasma of the temperature T. For the $He^4$ fusion products the range is given by

$$\lambda_0 = \frac{a(kT)^{3/2}}{n}, \qquad a = 2.5\times 10^{34} \text{ [cgs]} \tag{8}$$

This time then is

$$\tau = \frac{\lambda_0}{v} = \frac{a(kT)^{3/2}}{nv} \tag{9}$$

It has to be compared with the time the electron can escape the burning plasma behind the detonation front, given by

$$t_{esc} \approx r_0/v_e \tag{10}$$

where $r_o$ is the radius of the burning deuterium cylinder, and $v_e$ the electron velocity.
One thus has,

$$\frac{t_{esc}}{\tau} = \frac{v}{v_e}\frac{nr_0}{a(kT)^{3/2}} \tag{11}$$

Putting $n = 10^{23}$ cm$^{-3}$, $T \approx 10^8$ K with $v_e \sim 10^9$ cm/s, $v \sim 10^9$ cm/s one finds $t_{esc}/\tau \sim 2.5\, r_0$. Therefore, for $t_{esc} < \tau$ requires that $r_0 < 0.4$cm.

To reach the highest specific impossible, one should make the deuterium cylinder as thin as possible.

## 5. Magnetic entrapment of the charged fusion products and the stopping of the proton beam in dense deuterium

If a fission bomb is used to trigger a thermonuclear detonation, there is so much energy available that almost any radiation implosion configuration is likely going to work. As an example, one may place a fission bomb and a sphere of solid deuterium in a shell of gold, in the two foci of an ellipsoidal cavity. The radiation released by the exploding fission bomb, by ablating the gold, launches a convergent shock wave into the liquid deuterium. With the temperature in the shock wave approximately rising as $1/r$, where $r$ is the distance of the shock wave from the center of the deuterium sphere, the ignition temperature is reached at some distance from the center. But only if this distance is larger than the stopping length of the DD fusion reaction products, typically a few cm, is a radially outward moving detonation wave ignited. This configuration is essentially the same kind of "hohlraum" (cavity) configuration, used in the indirect drive mode of laser fusion for a small DT sphere.

A configuration of this kind can still be used to burn deuterium, if the DT micro-explosion is used to trigger a larger deuterium explosion. For a starship which shall depend on deuterium as its only rocket fuel, and nothing else, this possibility is excluded.

But there is another possibility. It arises if the ignition is done with a $10^7$ Ampere-GeV proton (or deuterium) beam. If focused onto the one end of a slender cylindrical deuterium rod, the beam not only can be made powerful enough to ignite the deuterium, but its strong azimuthal magnetic field entraps the charged DD reaction fusion products within the deuterium cylinder, launching a deuterium detonation wave propagating with supersonic speed down the cylinder [13]. There the fusion gain and yield can in principle be made arbitrarily large, only depending on the length of the deuterium rod.

The range of the charged fusion products is determined by their Larmor radius

$$r_l = \frac{\alpha}{B} \tag{12}$$

where,

$$\alpha = \frac{c}{e}\frac{(2MAE)^{1/2}}{Z} \tag{13}$$

In (13) $c$, $e$ are the velocity of light and the electron charge, $M$ the hydrogen mass, $A$ the atomic weight and $Z$ the atomic number. $E$ is the kinetic energy of the fusion products.

If the magnetic field is produced by the proton beam current $I$, one has at the surface of the deuterium cylinder the azimuthal magnetic field

$$B_{\phi} = 0.2\, I/r \tag{14}$$

Combining (12) with (14) and requesting that $r_l < r$, one finds that

$$I > I_c \tag{15}$$

where $I_c = 5\alpha$. In **table 2** the values for $\alpha$ and $I_c$ for all the charged fusion products of the DD reaction are compiled. For all of them the critical current is below $I_c = 3.84 \times 10^6$ A. Therefore, with the choice $I \sim 10^7$ A, all the charged fusion products are entrapped inside the deuterium cylinder.

| Reaction | Fusion Product | Energy [MeV] | α [G cm] | $I_c$ [A] |
|---|---|---|---|---|
| DT | $He^4$ | 3.6 | $2.7\times 10^5$ | $1.35\times 10^6$ |
| DD | $He^3$ | 0.8 | $1.12\times 10^5$ | $5.6\times 10^5$ |
| DD | T | 1.0 | $2.5\times 10^5$ | $1.25\times 10^6$ |
| DD | H | 3.0 | $2.5\times 10^5$ | $1.25\times 10^6$ |
| $DHe^3$ | H | 14.65 | $5.56\times 10^5$ | $3.84\times 10^6$ |
| $DHe^3$ | $He^4$ | 3.66 | $2.78\times 10^5$ | $1.39\times 10^6$ |

**Table 2**. Critical Ignition Currents for Thermonuclear Reactions.

For the argon ion laser configuration proposed for the launch into earth orbit, where a small amount of DT serves as a trigger for the ignition of a larger amount of deuterium, the ignition of a magnetic field supported detonation wave in deuterium is there possible with an auxiliary high explosive driven megampere current generator, setting up an axial magnetic field, by an azimuthal current around the rod. The charged fusion products are there spiraling down the rod. The current needed to entrap the charged fusion products are there of the same order of magnitude, that is of the order of $\sim 10^7$ A.

For the deuterium-tritium thermonuclear reaction the condition for propagating burn in a sphere of radius $r$ and density $\rho$, heated to a temperature of $10^8$ K, is given by $\rho r \geq 1$ g/cm$^2$. This requires energy of about 1 MJ. For the deuterium reaction this condition is $\rho r \geq 10$ g/cm$^2$, with an ignition temperature about 10 times larger. That a thermonuclear detonation in deuterium is possible at all is due to the secondary combustion of the T and $He^3$ DD fusion reaction products [8]. The energy required there would be about $10^4$ times larger or about $10^4$ MJ, for all practical purposes out of reach for non-fission ignition. However, if the ignition and burn is along a deuterium cylinder, where the charged fusion products are entrapped by a magnetic field within the cylinder, the condition $\rho r \geq 10$ g/cm$^2$ is replaced by

$$\rho z \geq 10 \ \text{gcm}^{-2} \tag{16}$$

where z is the length of the cylinder.

If the charged fusion products are entrapped within the deuterium cylinder, and if the condition $\rho z > 10$ g/cm$^2$ is satisfied, and finally, if the beam energy is large enough that a length $z$ > (10/$\rho$) cm of the cylinder is heated to a temperature of $10^9$ K, a thermonuclear detonation wave can propagate down the cylinder. This then leads to large fusion gains.

The stopping length of single GeV protons in dense deuterium is much too large to fulfill inequality (16). But this is different for an intense beam of protons, where the stopping length is determined by the electrostatic proton-deuteron two-stream instability [14]. In the presence of a strong azimuthal magnetic field the beam dissipation is enhanced by the formation of a collision less shock [15]. With the thickness of the shock by order of magnitude equal to the Larmor radius of the deuterium ions at a temperature of $10^9$ K, which for a magnetic field of the order $10^7$ G is of the order of $10^{-2}$ cm. For the two-stream instability alone, the stopping length is given by

$$\lambda \cong \frac{1.4c}{\varepsilon^{1/3}\omega_i} \tag{17}$$

Where $c$ is the velocity of light, and $\omega_i$ the proton ion plasma frequency, furthermore $\varepsilon = n_b/n$, with $n$ the deuterium target number density and $n_b$ = $2\times10^{16}$ cm$^{-3}$ the proton number density in the beam. For a 100-fold compressed deuterium rod one has $n = 5\times10^{24}$ cm$^{-3}$, with $\omega_i = 2\times10^{15}$ s$^{-1}$. One there finds that $\varepsilon = 4\times10^{-9}$ and, $\lambda \cong 1.2\times10^{-2}$ cm. This short length, together with the formation of the collision-less magneto-hydrodynamic shock, ensures the dissipation of the beam energy into a small volume at the end of the deuterium rod. For a deuterium number density $n = 5\times10^{24}$ cm$^{-3}$ one has $\rho = 17$ g/cm$^3$, and to have $\rho z > 10$ g/cm$^2$, then requires that $z \geq 0.6$ cm. With $\lambda < z$, the condition for the ignition of a thermonuclear detonation wave is satisfied. The ignition energy is given by

$$E_{ign} \sim 3nkT\pi r^2 z \tag{18}$$

where $T \approx 10^9$ K.

For 100-fold compressed deuterium, one has $\pi r^2 = 10^{-3}$ cm$^2$, when initially it was $\pi r^2 = 10^{-1}$ cm$^2$. With $\pi r^2 = 10^{-3}$ cm$^2$, $z = 0.6$ cm one finds that $E_{ign} \leq 10^{16}$ erg or ≤ 1GJ. This energy is provided by the $10^7$ Ampere-GeV proton beam lasting $10^{-7}$s. The time is short enough to ensure the cold compression of deuterium to high densities. For a $10^3$-fold compression, found feasible in laser fusion experiments, the ignition energy is ten times less.

In hitting the target, a fraction of the proton beam energy is dissipated into X-rays by entering and bombarding the high *Z* material cone, focusing the proton beam onto the deuterium cylinder. The X-rays released fill the hohlraum surrounding the deuterium cylinder, compressing it to high densities, while the bulk of the proton beam energy heats and ignites the deuterium cylinder at its end, launching in it a detonation wave.

If the GeV -$10^7$ Ampere proton beam passes through background hydrogen plasma with a particle number density *n*, it induces in the plasma a return current carried by its electrons, where the electrons move in the same direction as the protons. But because the current of the proton beam and the return current of the plasma electrons are in opposite directions, they repel each other. Since the stagnation pressure of the GeV -$10^7$ Ampere proton beam is much larger than the stagnation pressure of the electron return current, the return current electrons will be repelled from the proton beam towards the surface of the proton beam.

The stagnation pressure of a GeV proton beam is ($M_H$ proton mass)

$$p_i \cong \rho_i c^2 = n_b M_H c^2 \tag{19}$$

For $n_b \cong 2\times10^{16}$ cm$^{-3}$ one obtains $p_i \cong 3\times10^{13}$ dyn/cm$^3$. For the electron return current one has (*m* electron mass)

$$p_e = n_e m v^2 \tag{20}$$

With the return current condition $n_e e v_e = n_i e v_i$, where for GeV protons $v_i \cong c$, one has

$$v_e / c = n_i / n_e \tag{21}$$

Taking the value $n_e = 5\times10^{22}$ cm$^{-3}$, valid for uncompressed solid deuterium, one obtains $v_e \cong 10^4$ cm/s and hence $p_e \cong 5\times10^3$ dyn/cm$^2$. This is negligible against $p_i$, even if $n_e$ is $10^3$ times larger, as in highly compressed deuterium. The assumption, that the magnetic field of the proton beam is sufficiently strong to entrap the charged fusion products within the deuterium cylinder, is therefore well justified.

## 6. Solution in between two extremes

With chemical propulsion manned space flight to the moon is barely possible and only with massive multistage rockets. For manned space flight beyond the moon, nuclear propulsion is indispensible. Nuclear thermal propulsion is really not much better than advanced chemical propulsion. Ion propulsion, using a nuclear reactor driving an electric generator has a much higher specific impulse, but not enough thrust for short interplanetary transit times, as they are needed for manned missions. This leaves the propulsion by a chain of fission bombs (or fission triggered fusion bombs) as the only credible option. There the thrust and specific impulse are huge in comparison. But a comparatively small explosive yield is there be desirable. Making the yield too small, the bombs become extravagant in the sense that only a small fraction of the fission explosive is consumed. The way to overcome this problem is the non-fission ignition of small fusion explosions. A first step in this direction is the non-fission ignition of deuterium-tritium (DT) thermonuclear micro-explosions; the easiest one to be ignited, expected to be realized in the near future. Because of it, I had chosen this reaction for the first proposed thermonuclear micro-explosion propulsion concept, with the ignition done by an intense relativistic electron beam [1, 2]. But because in the DT reaction 80% of the energy is released into neutrons which cannot be reflected from the spacecraft by a magnetic mirror, it was proposed to surround the micro-explosion with a neutron-absorbing hydrogen propellant, increasing the thrust on the expense of the specific impulse. It was for this reason that in the "Daedalus" interstellar probe study of the British Interplanetary Society [3], the neutron-less helium3-deuterium ($He^3$-D) reaction was proposed, because for such a mission the specific impulse should be as high as possible. But even in a $He^3$-D plasma, there are a some neutron producing DD reactions. There is no large source of $He^3$ on the Earth, even though it might exist on the surface of the moon, and in the atmosphere of Jupiter. In the DD reaction much less energy goes into neutrons, but it is more difficult to ignite. The situation is illustrated in Fig 1. On the left side it shows the experimentally verified ignition of a DT pellet with the X-rays drawn in an underground test from a fission bomb (Centurion Halite experiment at the Nevada Test Site). For the ignition of the DT reaction with propagation thermonuclear burn (i.e. detonation) requires a few megajoule, with a density$\times$radius target product, $\rho r$ >1g/cm$^2$. And on the right side is the 15 Megaton "Mike" test, where with the Teller-Ulam configuration a large amount of liquid deuterium is ignited with a fission bomb. For the DD reaction propagating thermonuclear burn (i.e. detonation) requires that $\rho r$ $\geq$10g/cm$^2$. In between is the proposed hypothetical deuterium target, where a detonation wave in a thin cylindrical deuterium rod is ignited by a pulsed $10^7$ Ampere-GeV proton beam, utilizing the strong magnetic field of the beam current.

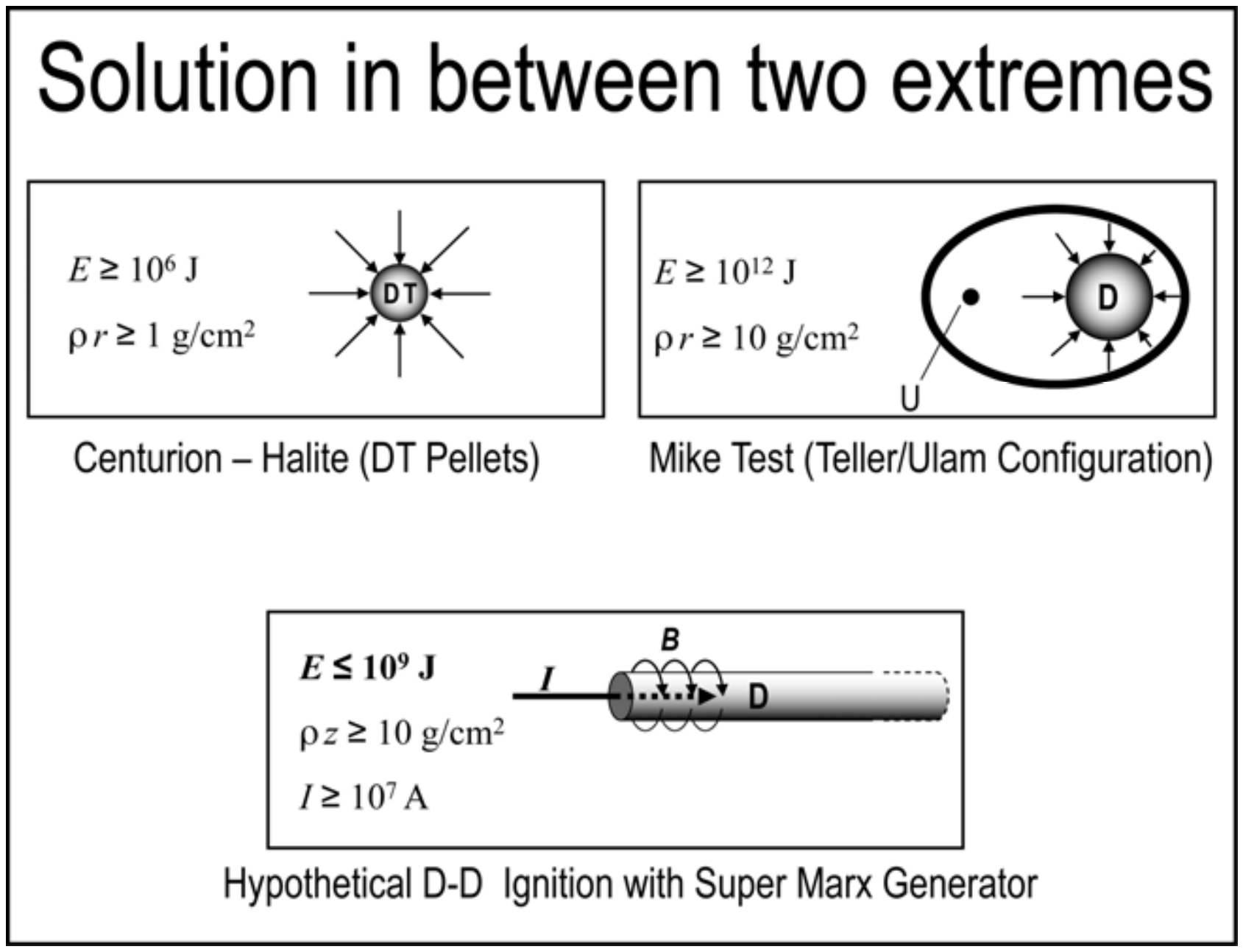

**Figure 1** Ignition with $10^7$ Ampere GeV proton beam [16].

To estimate the order of magnitude what is needed, we consider a spacecraft with a mass of $M_0 = 10^3$ ton $= 10^9$ g, to be accelerated by one $g \cong 10^3$ cm/s$^2$, with a thrust $T = M_0 g \cong 10^{12}$ dyn. To establish the magnitude and number of fusion explosions needed to propel the spacecraft to a velocity of $v = 100$ km/s = $10^7$ cm/s, we use the equation

$$T = c\frac{dm}{dt} \tag{22}$$

where we set $c \cong 10^8$ cm/s, equal to the expansion velocity of the fusion bomb plasma.
We thus have

$$\frac{dm}{dt} = \frac{T}{c} = 10^4\ g/s = 10\,kg/s \tag{23}$$

The propulsion power is given by

$$P = \frac{c^2}{2}\frac{dm}{dt} = \frac{c}{2}T \tag{24}$$

in our example it is $P = 5\times10^{19}$ erg/s.

With $E = 4\times10^{19}$ erg equivalent to the explosive energy of one kiloton of TNT, $P$ is equivalent to about one nuclear kiloton bomb per second.
From the rocket equation

$$v = c\ln\left(\frac{M_0}{M_1}\right) \tag{25}$$

where $M_0$ is the initial, and $M_1$ the final mass, and setting $M_1 = M_0 - \Delta M$, where $\Delta M << M_0$ is the mass of all used up bombs one has,

$$v \cong c\ln\left(1+\frac{\Delta M}{M_0}\right) \cong c\frac{\Delta M}{M_0} \tag{26}$$

where $v$ is the velocity reached by the spacecraft after having used up all the bombs of mass $\Delta M$. If one bomb explodes per second, its mass according to (23) is $m_0 = 10^4$ g.

Assuming that the spacecraft reaches a velocity of $v = 100$ km/s $= 10^7$ cm/s, the velocity needed for fast interplanetary travel, one has $\Delta M = 10^8$ g, requiring $N = \Delta M/m_0 = 10^4$ one kiloton fusion bombs, releasing the energy $E_b = 5\times10^{19}\times10^4 = 5\times10^{23}$ erg. By comparison, the kinetic energy of the spacecraft $E_s = (1/2)M_0 v^2 = 5\times10^{22}$ erg, is 10 times less. In reality it is still smaller, because a large fraction of the energy released by the bomb explosions is dissipated into space.

One can summarize these estimates by concluding that a very large number of nuclear explosions is needed, which for fission explosions, but also for deuterium-tritium explosions, would become very expensive. This strongly favors deuterium, more difficult to ignite in comparison to a mixture of deuterium with tritium, but abundantly available. Here I will try to show how bomb propulsion solely with deuterium might be possible.

## 7. The non-fission ignition of small deuterium nuclear explosives

With no deuterium-tritium (DT) micro-explosions yet ignited, the non-fission ignition of pure deuterium (DD) fusion explosions seems to be a tall order. An indirect way to reach this goal is by staging a smaller DT explosion with a larger DD explosion. There the driver energy, but not the driver may be rather small. A direct way requires a driver with order of magnitude larger energies.

I claim that the generation of GeV potential wells, made possible with magnetic insulation of conductors levitated in ultrahigh vacuum (in a laboratory on Earth), has the potential to lead to order of magnitude larger driver energies [1]. It is the ultrahigh vacuum of space by which this can be achieved without levitation. Therefore, the spacecraft acting as a capacitor can be charged up to GeV potentials.

If charged to a positive GeV potential, a gigajoule intense relativistic ion beam below the Alfvén current limit can be released from the spacecraft and directed to the deuterium explosive for its ignition. If the current needed for ignition is below the Alfvén limit for ions, the beam is "stiff". The critical Alfvén current for protons is $I_A = 3.1\times10^7 \beta\gamma$ [A], where $\beta = v/c$, $\gamma = \left(1-\beta^2\right)^{-1/2}$, with $v$ the proton velocity and $c$ the velocity of light. For GeV protons $I_A$ is well in excess of the critical current (15) to entrap the DD fusion reaction products, the condition for detonation [13].

In a possible bomb configuration shown in Fig.2, the liquid (or solid) D explosive has the shape of a long cylinder, placed inside a cylindrical "hohlraum" **h**. A GeV proton beam **I** coming from the left, in entering the hohlraum dissipates part of its energy into a burst of X-rays compressing and igniting the D bomb-cylinder. With its gigajoule energy lasting less than $10^{-7}$ s, the beam power is greater than $10^{16}$ Watt, sufficiently large to ignite the D explosive. The main portion of the beam energy is focused by the cone onto the deuterium rod, igniting at its end a detonation wave.

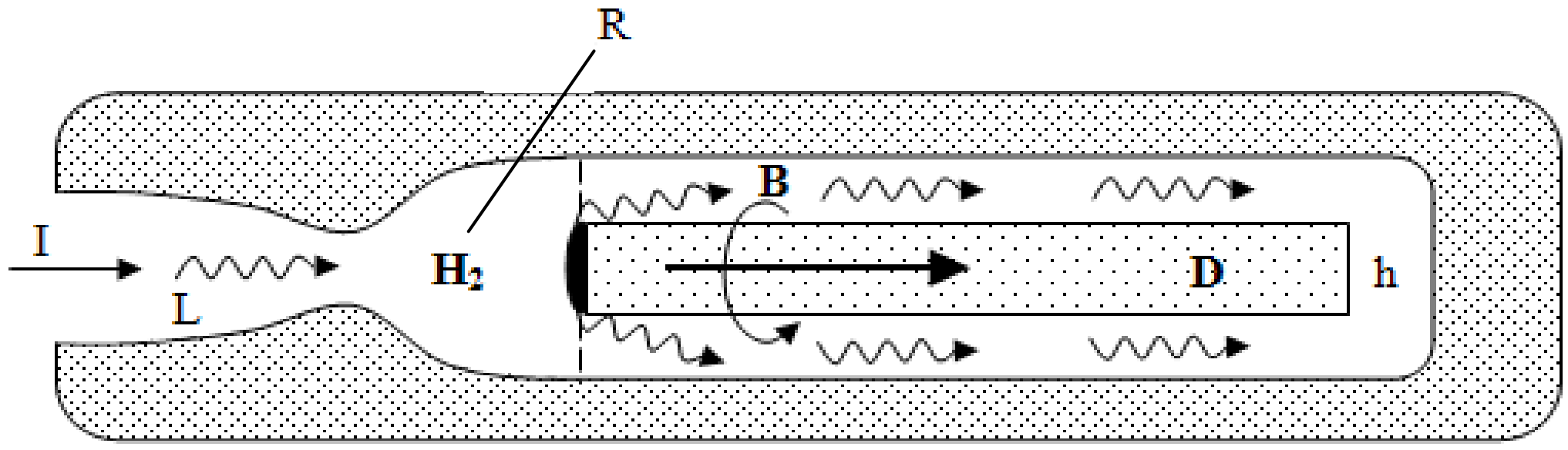

**Figure 2** Pure deuterium fusion explosion ignited with an intense ion beam. **D** deuterium rod, **h** hohlraum, **I** ion beam, **B** magnetic field, **R** miniature target rocket chamber, **$H_2$** solid hydrogen, **L** laser beam to heat hydrogen in miniature rocket chamber.

Because the condition for thermonuclear burn and detonation depends only on the critical current (15), but not on the radius of the deuterium cylinder, one may wish to make the diameter of the deuterium cylinder as small as possible, because as it was shown above, the specific impulse can become largest, with the electrons not taken away kinetic energy from the charged fusion products. There then, the yield of the deuterium fusion explosions can be made much smaller, eliminating the need for Orion-type shock absorbers, but the thrust there is also much smaller. For very deep space missions that would not be a disadvantage.

A problem in either case is that 38% of the energy is released as neutrons. In hitting the space craft they lead to its heating, requiring a presumably large radiator. For a long and thin deuterium rod this problem can likely be reduced by a boron diaphragm on the deuterium rod, as shown in **Fig 3**, because boron is a good neutron absorber. It can be enhanced by a hydrogen moderator, because the absorption cross section for neutrons is greatly increased with a reduced neutron kinetic energy. Both the boron and the hydrogen there simply become part of the propellant, reducing the specific impulse but increasing the thrust.

We know that in comets there is large amount of deuterium, readily available for mining. And we know that comets have also nitrogen and carbon. From this knowledge it is very likely, that other light elements like boron should be in relative high concentrations.

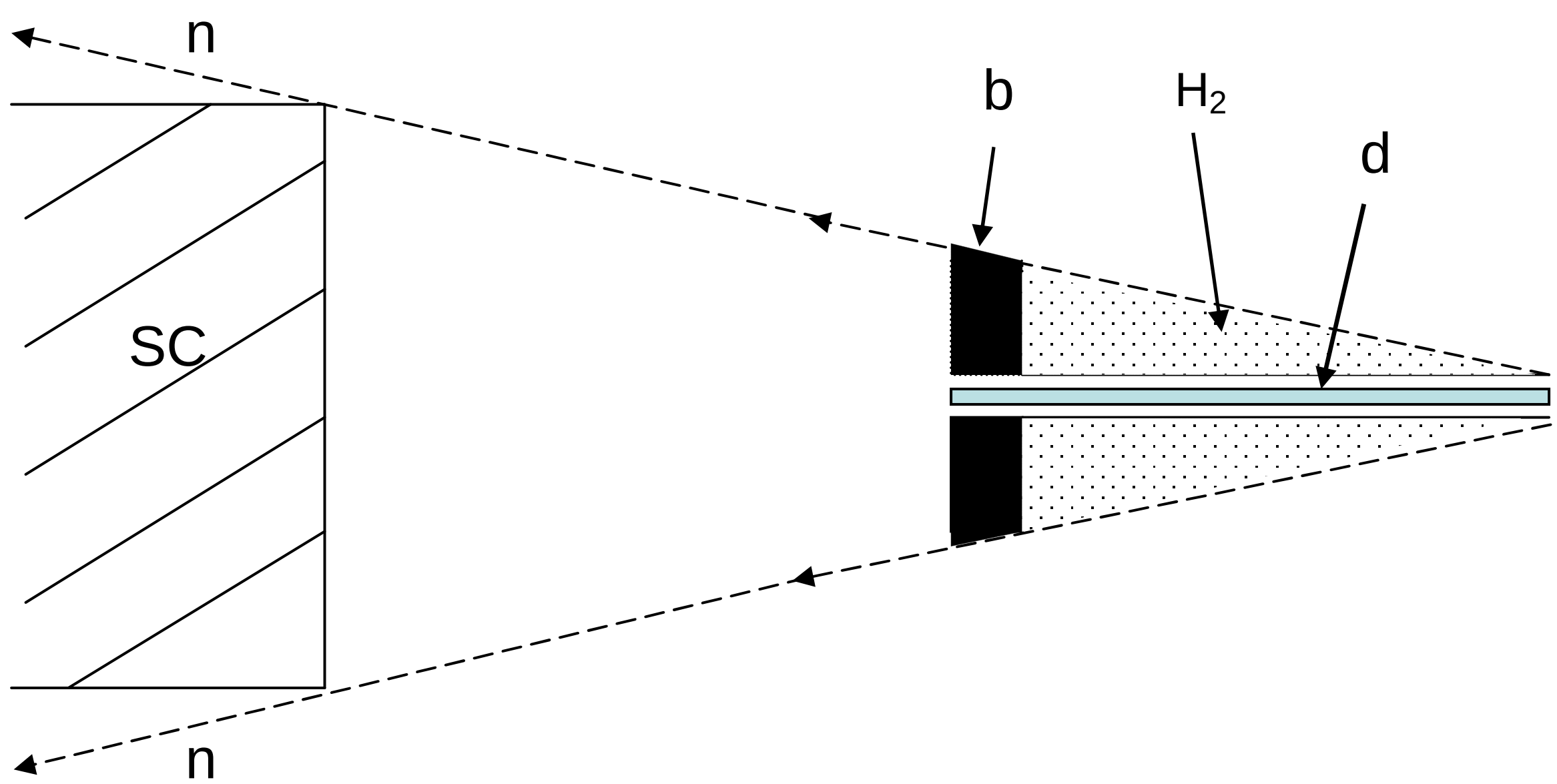

**Figure 3:** Screening of the space craft SC against the neutrons, n, by a boron diaphragm, b, with solid hydrogen $H_2$ as a neutron moderator to increase the neutron absorption cross section of boron, with the neutrons released from deuterium cylinder d.

The problem of the waste heat radiator remains high, but it favors large explosions because there most of the waste heat goes with the propellant into space. Droplet radiators, with the droplets slowly evaporating, are unlikely to work. Placing the neutron absorbing radiators near the shock absorber, permitting them to get red-hot, and thermally insulating the rest of the space craft from the radiators, may solve the problem.

## 8. Delivery of a Gev proton beam onto the deuterium fusion explosive

The spacecraft is inductively charged against an electron cloud surrounding the craft, and, with a magnetic field of the order $10^4$ G, easily reached by superconducting currents flowing in an azimuthal direction around the craft, is magnetically insulated against the electron cloud up to GeV potentials. The spacecraft and its surrounding electron cloud form a virtual diode with a GeV potential difference. To generate a proton beam, it is proposed to attach a miniature hydrogen filled rocket chamber **R** to the deuterium bomb target, at the position where the proton beam hits the fusion explosive (see **Fig. 2**). A pulsed laser beam from the spacecraft is shot into the rocket chamber, vaporizing the hydrogen, which is emitted through the Laval nozzle as a supersonic plasma jet. If the nozzle is directed towards the spacecraft, a conducting bridge is established, rich in protons between the spacecraft and the fusion explosive. Protons in this bridge are then accelerated to GeV energies, hitting the deuterium explosive. Because of the large dimension of the spacecraft, the jet doesn't have to be aimed at the spacecraft very accurately.

The original idea for the electrostatic energy storage on a magnetically insulated conductor was to charge up a levitated superconducting ring to GeV potentials, with the ring magnetically insulated against breakdown by the magnetic field of a large toroidal current flowing through the ring. It is here proposed to give the spacecraft a topologically equivalent shape, using the entire spacecraft for the electrostatic energy storage (see **Fig. 3**). There, toroidal currents flowing azimuthally around the outer shell of the spacecraft, not only magnetically insulate the spacecraft against the surrounding electron cloud, but the currents also generate a magnetic mirror field which can reflect the plasma of the exploding fusion bomb. In addition, the expanding bomb plasma can induce large currents, and if these currents are directed to flow through magnetic field coils positioned on the upper side of the spacecraft, electrons from there can be emitted into space surrounding the spacecraft by thermionic emitters placed on the inner side of these coils, inductively charging [12] the spacecraft for subsequent proton beam ignition pulses. A small high voltage generator driven by a small onboard fission reactor can make the initial charging, ejecting from the spacecraft negatively charged pellets.

With the magnetic insulation criterion, $E < B$ ($E, B,$ in electrostatic units, esu), where $B$ is the magnetic field surrounding the spacecraft measured in Gauss, then for $B \cong 10^4$G, $E = 3 \times 10^3$ esu $= 9 \times 10^5$ V/cm, one has $E \sim (1/3)B$ hence $E < B$. A spacecraft with the dimension $l \sim 3\times10^3$ cm, can then be charged to a potential $El \sim 3\times10^9$ Volts, with the stored electrostatic energy is of the order $\varepsilon \sim \left(E^2/8\pi\right)\, l^3$.

For $E = 3\times10^3$ esu, and $l = 3\times10^3$ cm, $\varepsilon$ is of the order of one gigajoule. The discharge time is of the order $\tau \sim l/c$, where $c = 3\times10^{10}$ cm/s is the velocity of light. In our example we have $\tau \sim 10^{-7}$ sec. For a proton energy pulse of one gigajoule, the beam power is $3\times10^{16}$ erg/s = 30 petawatt, large enough to ignite a pure deuterium explosion.

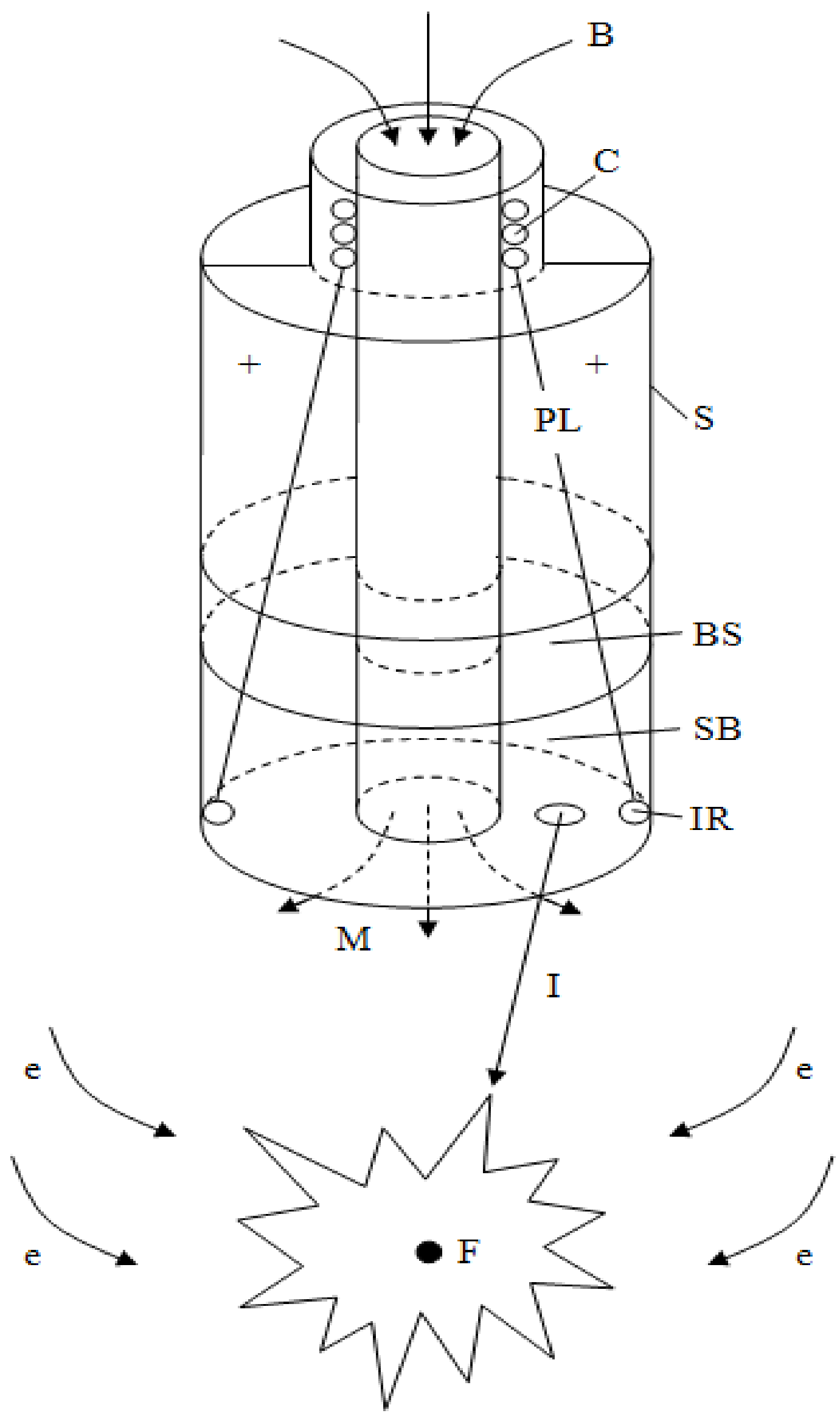

**Figure 4** Superconducting "atomic" spaceship, positively charged to GeV potential, with azimuthal currents and magnetic mirror **M** by magnetic field **B**. **F** fusion minibomb in position to be ignited by intense ion beam **I**, **SB** storage space for the bombs, **BS** bioshield for the payload **PL**, **C** coils pulsed by current drawn from induction ring **IR**. **e** electron flow neutralizing space charge of the fusion explosion plasma.

## 9. Lifting of large payloads into earth orbit

To lift large payloads into earth orbit remains the most difficult task. For a launch from the surface of the earth, magnetic insulation inside the earth atmosphere fails, and with it the proposed pure deuterium bomb configuration. A different technique is here suggested, which I had first proposed in a classified report, dated January of 1970 [9], declassified July 2007, and thereafter published [17]. A similar idea was proposed in a classified Los Alamos report, dated November 1970 [18], declassified July 1979. In both cases the idea is to use a replaceable laser for the ignition of each nuclear explosion, with the laser material thereafter becoming part of the propellant. The Los Alamos scientists had proposed to use an infrared carbon dioxide ($CO_2$) or chemical laser for this purpose, but this idea does not work, because the wavelength is too long, and therefore unsuitable for inertial confinement fusion. I had suggested an ultraviolet argon ion laser instead. However, since argon ion lasers driven by an electric discharge have a small efficiency, I had suggested a quite different way for its pumping, illustrated in **Fig. 5**. There the efficiency can be expected to be quite high. It was proposed to use a cylinder of solid argon, surrounding it by a thick cylindrical shell of high explosive. If simultaneously detonated from outside, a convergent cylindrical shockwave is launched into the argon. For the high explosive one may choose hexogen with a detonation velocity of 8 km/s. In a convergent cylindrical shockwave the temperature rises as $r^{-0.4}$, where $r$ is the distance from axis of the cylindrical argon rod. If the shock is launched from a distance of ~1 m onto an argon rod with a radius equal to 10 cm, the temperature reaches 90,000 K, just right to excite the upper laser level of argon. Following its heating to 90,000 K the argon cylinder radially expands and cools, with the upper laser level frozen into the argon. This is similar as in a gas dynamic laser, where the upper laser level is frozen in the gas during its isentropic expansion in a Laval nozzle. To reduce depopulation of the upper laser level during the expansion by super-radiance, one may dope to the argon with a saturable absorber, acting as an “antiknock” additive. In this way megajoule laser pulses can be released within 10 nanoseconds. A laser pulse from a small Q-switched argon ion laser placed in the spacecraft can then launch a photon avalanche in the argon rod, igniting a DT micro-explosion.

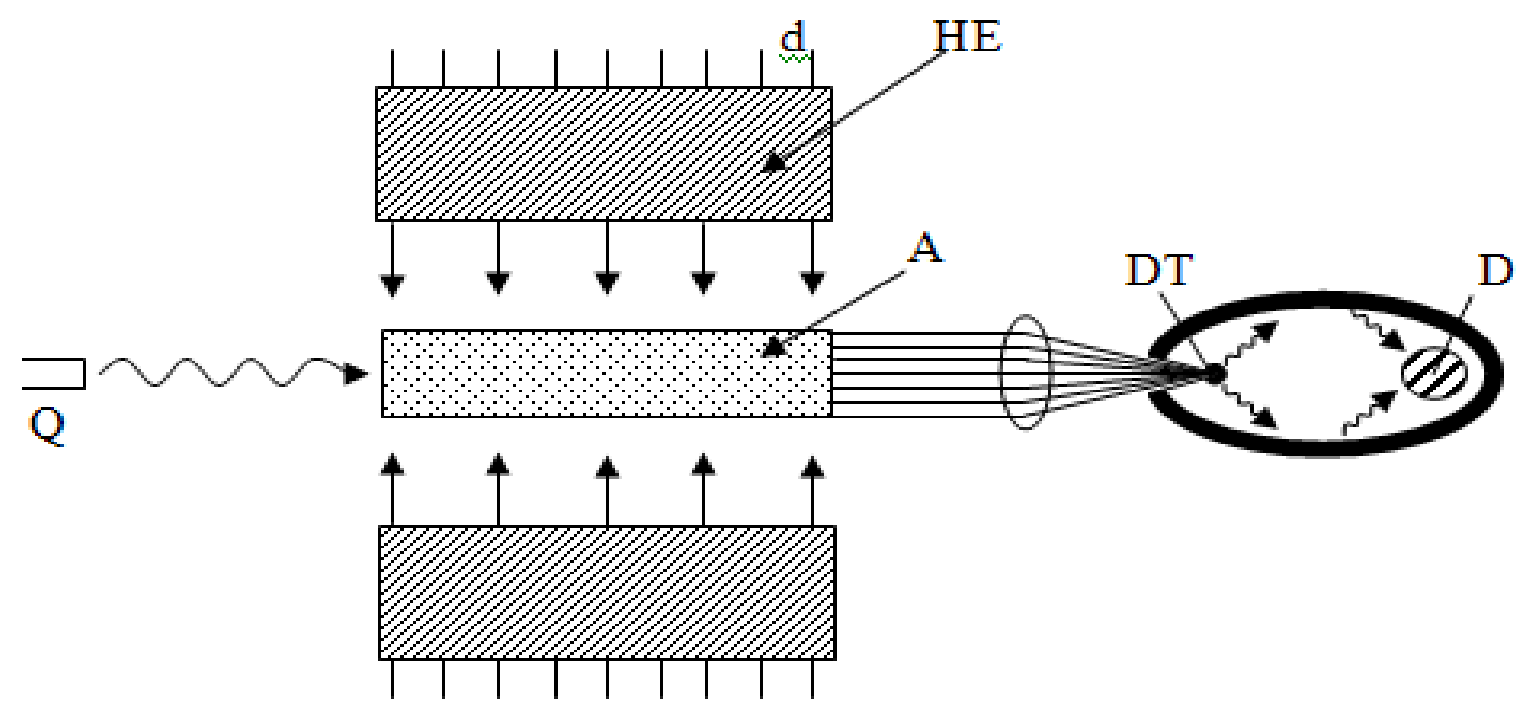

**Figure 5** Argon ion laser igniter, to ignite a staged DT →DD fusion explosion in a mini-Teller-Ulam configuration. **A** solid argon rod. **HE** cylindrical shell of high explosive, **d** detonators. **Q** Q-switched argon ion laser oscillator.

Employing the Teller-Ulam configuration, by replacing the fission explosive with a DT micro-explosion, one can then ignite a much larger DD explosion.

As an alternative one may generate a high current linear pinch discharge with a high explosive driven magnetic flux compression generator. If the current $I$ is of the order $I = 10^7$A, the laser can ignite a DT thermonuclear detonation wave propagating down the high current discharge channel, which in turn can ignite a much larger pure DD explosion.

If launched from the surface of the earth, one has to take into account the mass of the air entrained in the fireball. The situation resembles a hot gas driven gun, albeit one of rather poor efficiency. There the velocity gained by the craft with $N$ explosions, each setting off the energy $E_b$, is given by

$$v = \sqrt{2N E_b / M_0} \tag{27}$$

For $E_b = 5\times 10^{19}$erg, $M_0 = 10^9$g, and setting for $v$ = 10 km/s = $10^6$ cm/s the escape velocity from the Earth, one finds that $N \geq 10$. Assuming an efficiency of 10%, about 100 kiloton explosions would there be needed.

## 10. Neutron entrapment in an autocatalytic thermonuclear detonation wave – a means to increase the specific impulse and to solve the large radiator problem

The principal reason why neutrons released by thermonuclear reactions pose such a serious problem is that they cannot be repelled from the spacecraft by a magnetic field. Choosing a neutron absorbing target as shown in **Fig. 3** one can though reduce the flux of neutrons hitting the spacecraft. Besides the material damage the neutrons can inflict on the spacecraft, it is that they release heat which must be removed by a radiator, and this radiator will be very large.

The idea of the autocatalytic thermonuclear detonation wave [13] presents a solution, which if feasible would very much reduce the magnitude of this problem. For its implementation it requires very large bremsstrahlungs flux densities in the burn zone behind the thermonuclear detonation front. Such large bremsstrahlungs flux densities will occur in deuterium detonation burn, at the highest temperature for all the thermonuclear reactions.

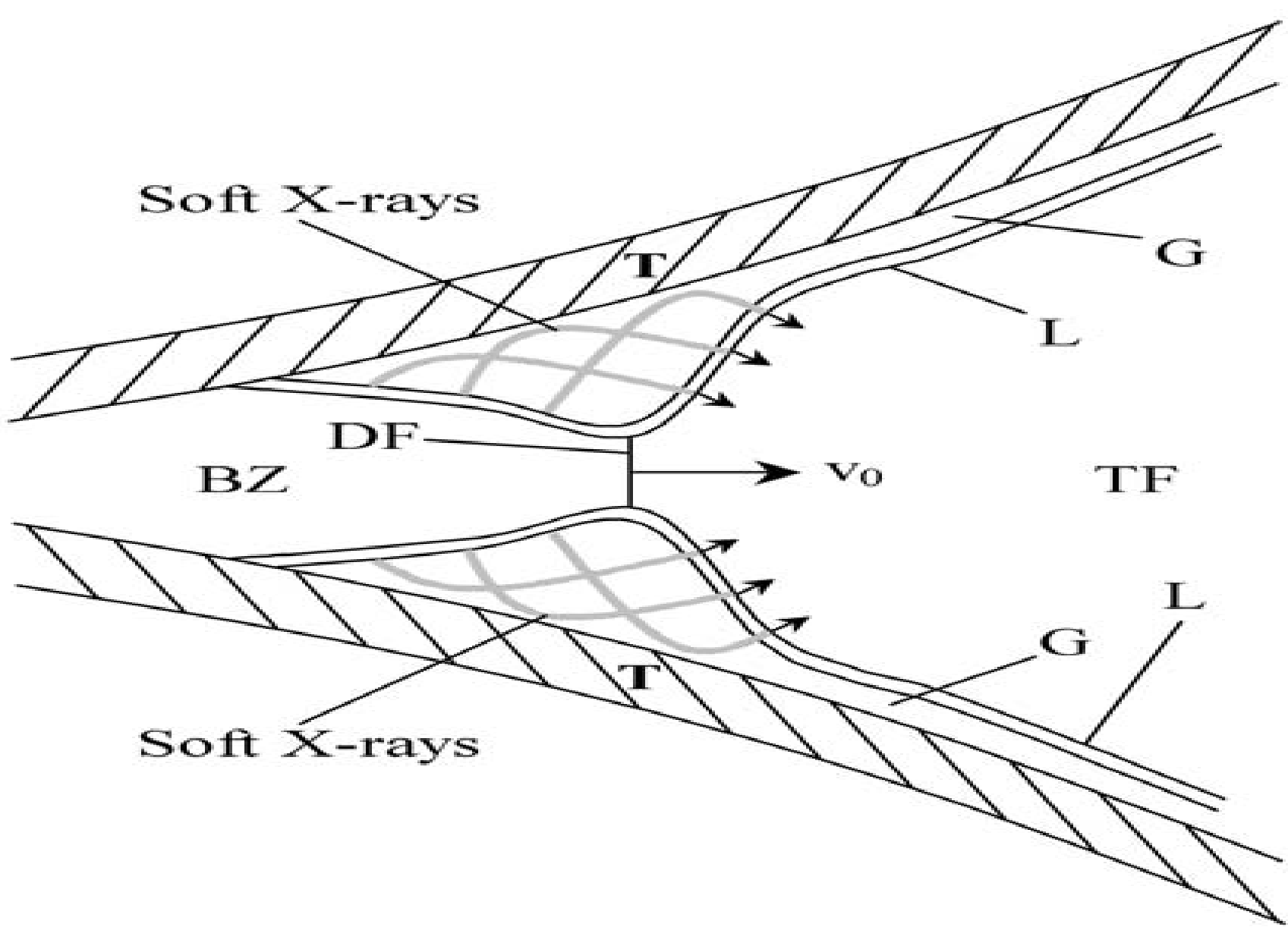

**Figure 6.** Autocatalytic thermonuclear detonation using a soft X-ray precursor from the burn zone BZ to precompress the thermonuclear fuel TF ahead of the detonation front DF. The soft x-rays travel through the gap G between the tamp T and the liner L.

In an Autocatalytic thermonuclear detonation, explained in **Fig. 6**, soft X-rays generated through the burn of the thermonuclear plasma behind the detonation front, compresses the still unburned thermonuclear fuel ahead of the front. The increase in the fuel density, both in the Teller-Ulam configuration and the autocatalytic thermonuclear detonation wave, is of crucial importance, with the reaction rate going in proportion to the square of the density.

From the burning plasma behind the detonation front, energy flows into all spatial directions. Part is by bremsstrahlung and part by electronic heat conduction. Roughly half of the energy flows into the liner, and a quarter into the still unburned fuel ahead of the wave and one quarter into opposite direction. The bremsstrahlung emission rate is given by

$$\varepsilon_r = 1.42 \times 10^{-27} n^2 \sqrt{T} \ \text{[erg/cm}^3\text{s]} \tag{28}$$

For T = $10^9$ K one has $\varepsilon_r \approx 3.2 \times 10^{-23}\, n^2$ [erg/cm$^3$s]. The flux of the bremsstrahlung which goes into the liner is $\varepsilon_r r/2$ , where r is the radius of the deuterium rod just behind the detonation front. About ½ of this radiation runs ahead of the detonation front where it pre-compresses the deuterium**.** Its intensity is:

$$\frac{\varepsilon_r r}{4} \approx 10^{-23} n^2 r \ \text{[erg/cm}^2\text{s]} \tag{29}$$

with $nr \approx 3 \times 10^{24}$cm$^{-2}$, (corresponding $\rho r \approx 10$ g/cm$^2$), one has

$$\frac{\varepsilon_r r}{4} \approx 30n \,\text{[erg/cm}^2\text{s]} \tag{30}$$

We are aiming at a density where the neutrons are absorbed in the burning plasma cylinder of radius *r*. If the neutron-deuteron collision cross section is $\sigma$ , then the neutron path length $\lambda_n$ must be smaller than *r*:

$$\lambda_n = \frac{1}{n\sigma} \le r \tag{31}$$

or

$$nr \ge \frac{1}{\sigma} \tag{32}$$

One typically has $\sigma \approx 10^{-24}$ cm$^2$ , hence

$$nr \ge 10^{24} \ \text{cm}^2 \tag{33}$$

If $r = 0.01$ cm then $n \geq 10^{26}$ cm$^{-3}$ = $5 \times 10^3\, n_o$ , where $n_o = 5 \times 10^{22}$ cm$^{-3}$, the particle number density of liquid deuterium. With $n$=$10^{26}$ cm$^{-3}$ , one finds that $\varepsilon_r r/4 \approx 3 \times 10^{27}$ erg/cm$^2$s. The flux on a deuterium tube of radius *r* and length *z,* where for the burn zone $r \approx z$, one obtains onto its surface $(\varepsilon_r/4) 2\pi r z \approx 10^{24}$ erg/s = $10^{17}$ watt = 100 petawatt, certainly powerful enough to compress the deuterium to more than 1000 fold density.

Through the entrapment of the neutrons goes an increase of the specific impulse. With 38% of the energy going into the kinetic energy of the neutrons, and 62% into charged fusion products, the specific impulse increased by the factor $\sqrt{1+38/62} = \sqrt{10/6.2} = 1.275$ This increases the maximum exhaust velocity from $v$=$1.5\times 10^9$ cm/s to $v$=$1.9\times 10^9$ cm/s = 0.063$c$.

In the course of their thermalization in the supercompressed plasma, the neutron absorption cross section is greatly increased, both in the liner and tamp, with the liner also compressed to high densities. If the liner and tamp are made from boron which has a large neutron absorption cross section, the space craft is only heated by a greatly reduced thermal neutron flux, and only this much smaller amount of heat must be removed by a radiator.

## 11. Testing the deuterium micro-detonation concept

For the Orion bomb propulsion concept testing was a serious problem. If tested on the earth it would have resulted in the large fallout of fission products. These tests would have been needed to study the survival of the pusher plate under the repeated exposure of kiloton fission explosions (or fusion boosted explosions). The tests could, of course, been carried out in space, but this would have been extremely expensive, because it would have required to launch by chemical rockets the huge Orion space ship into space. For the deuterium micro-detonation propulsion concept this is fortunately not necessary, because there exists an alternative way to generate a $10^7$ Ampere GeV proton beam, replacing the huge space ship used as a large capacitor to be charged up to gigavolt potentials, by a "Super Marx generator" [16]. This Super-Marx generator can achieve the same on earth what the huge spaceship can achieve in space: The generation of gigavolt – $10^7$ Ampere proton beams. Since the realization of pure deuterium beams would be obviously a breakthrough in fusion, the expenditure for the development of a super Marx Generator would be well justified.

Up until now nuclear fusion by inertial confinement has only been achieved using large fission explosives as a means (driver) for ignition. From this experience we know that the ignition is easy with sufficiently large driver energies, difficult to duplicate with lasers or electric pulse power by ordinary Marx generator. The problem therefore is not the configuration of the thermonuclear explosive, but the driver, be it for the ignition of pure deuterium (D) as in the Mike Test, or for the ignition of deuterium-tritium (DT), as in the Centurion-Halite experiment, because for sufficiently large driver energies the target configuration is of secondary importance.

I claim that substantially larger driver energies can be reached with the "Super Marx Generator". It can be viewed as a two-stage Marx generator, where a bank of ordinary Marx generators assumes the role of a first stage. If the goal is the much more difficult ignition of a pure deuterium micro-explosion, the Super Marx generator must in addition to deliver a much larger amount of energy (compared to the energy of the most powerful lasers), also generate a magnetic field in the thermonuclear target that is strong enough to entrap the charged DD fusion products within the target. Only then is the condition for propagating thermonuclear burn fulfilled.

**Fig. 7** shows the circuit of an ordinary Marx generator, and **Fig. 8** that of a Super Marx. An artist's view of a mile-long Super Marx generator, charged up by about 100 ordinary Marx generators, and connected to the chamber in which the thermonuclear target is placed, is shown in **Figs. (9-11)**.

As shown in **Fig. 12**, the Super Marx is made up of a chain of co-axial capacitors with a dielectric which has to withstand the potential difference of $10^7$ Volt between the inner and outer conductor.

Following their charging up the Super Marx generator, the Marx generators are disconnected from the Super Marx. If the capacitors of the Super Marx can hold their charge long enough, this can be done by mechanical switches.

To erect the Super Marx, its capacitors C are switched into series by circular spark gap switches S. The capacitors of the Super Marx are magnetically levitated inside an evacuated tunnel, and magnetically insulated against the wall of the tunnel by an axial magnetic field $B$, generated by super conducting external magnetic field coils M. The magnetic insulation criterion requires that $B > E$, where $B$ is measures in Gauss and $E$ in electrostatic cgs units. If $B = 3\times 10^4$ G for example, magnetic insulation is possible up to $E = 3\times 10^4$ esu $\approx 10^7$ V/cm, at the limit of electron field emission. To withstand a voltage of $10^9$ Volt between the outer positively charged surface of the capacitors in series and the tunnel wall, then requires a distance somewhat more than one meter.

The capacitance of one co-axial capacitor with the inner and outer radius, $R_0$ and $R_1$ of length $l$ and filled with a dielectric of dielectric constant $\varepsilon$ is

$$C = \varepsilon \frac{l}{2\ln(R_1/R_0)} \text{ [cm]} \tag{34}$$

Assuming a breakdown strength of the dielectric larger than $3\times 10^4$ V/cm, and a potential difference of $10^7$ Volt between the inner and outer conductor, the smallest distance of separation $d$ between both conductors has to be $d = R_1 - R_0 \cong 3\times 10^2$ cm. If for example $l = 1.6\times 10^3$ cm, $R_1 = l/2 = 8\times 10^2$ cm, and $\varepsilon \cong 10$, one finds that $C \cong 2\times 10^4$ cm. For these numbers the energy $e$ stored in the capacitor ($V = 10^7$ Volt $\cong 3\times 10^4$ esu) is

$$e = (1/2)\, CV^2 \cong 10^{13} \text{ erg} \tag{35}$$

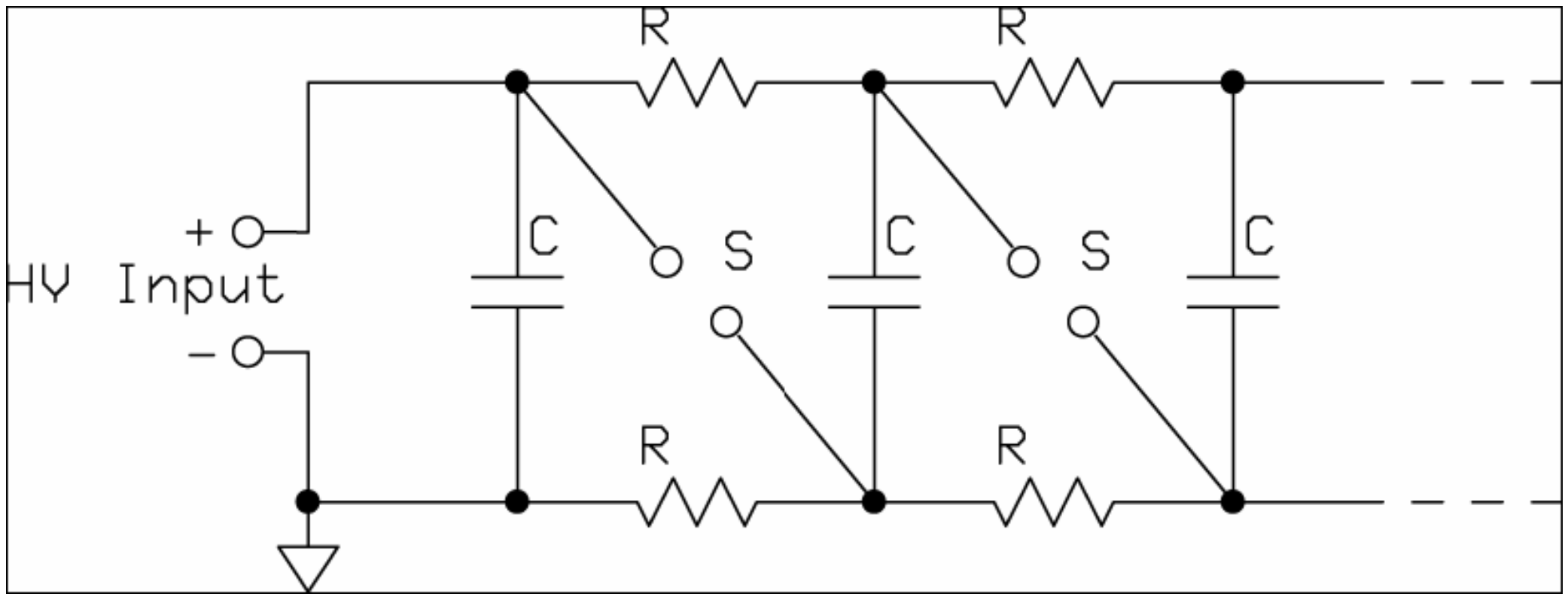

**Figure 7**: In an "ordinary" Marx generator n capacitors *C* charged up to the voltage *v*, and are over spark gaps switched into series, adding up their voltages to the voltage $V = nv$.

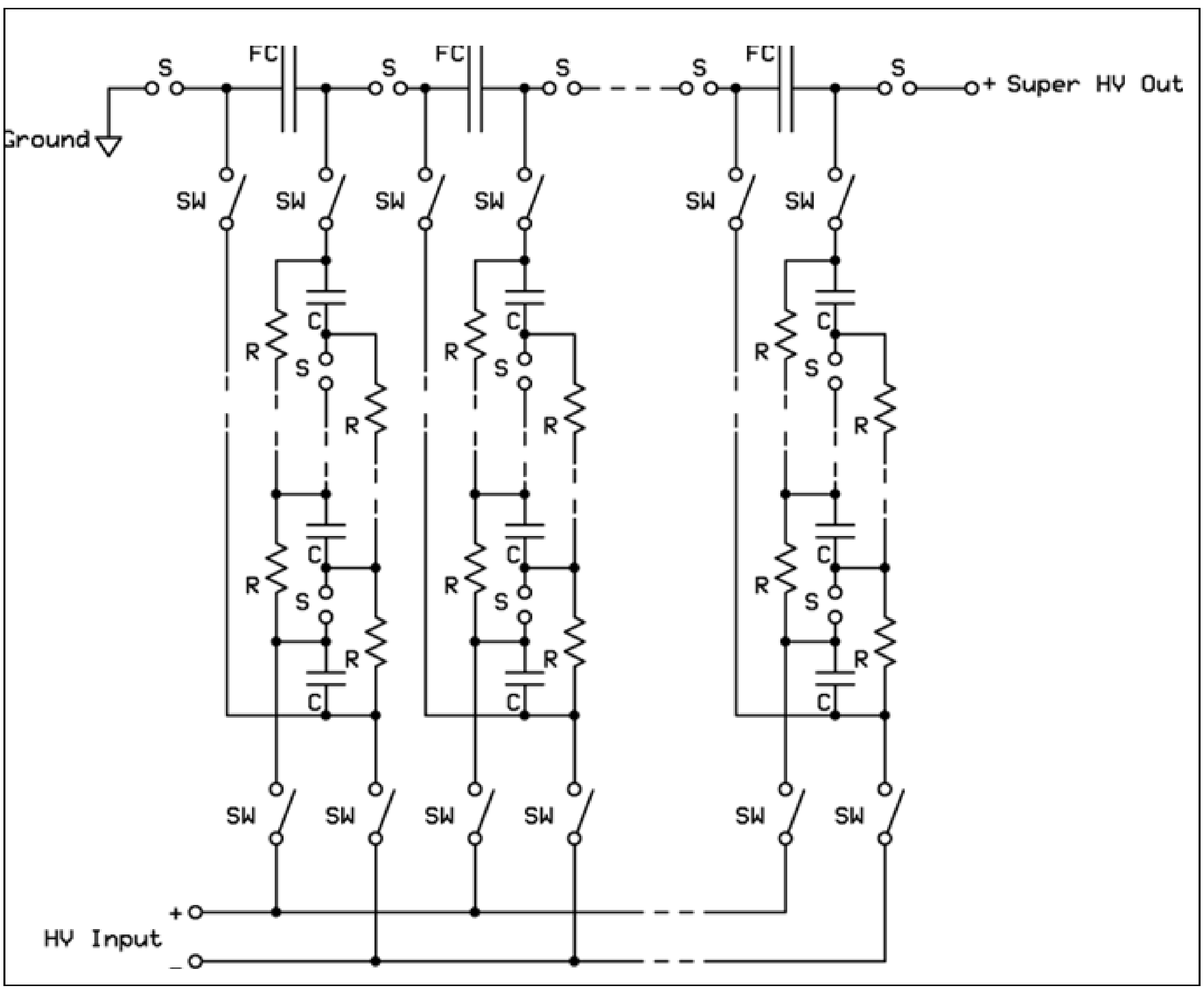

**Figure 8**: In a Super Marx generator, *N* Marx generators charge up *N* fast capacitors FC to the voltage *V*, which switched into series add up their voltages to the voltage *NV*.

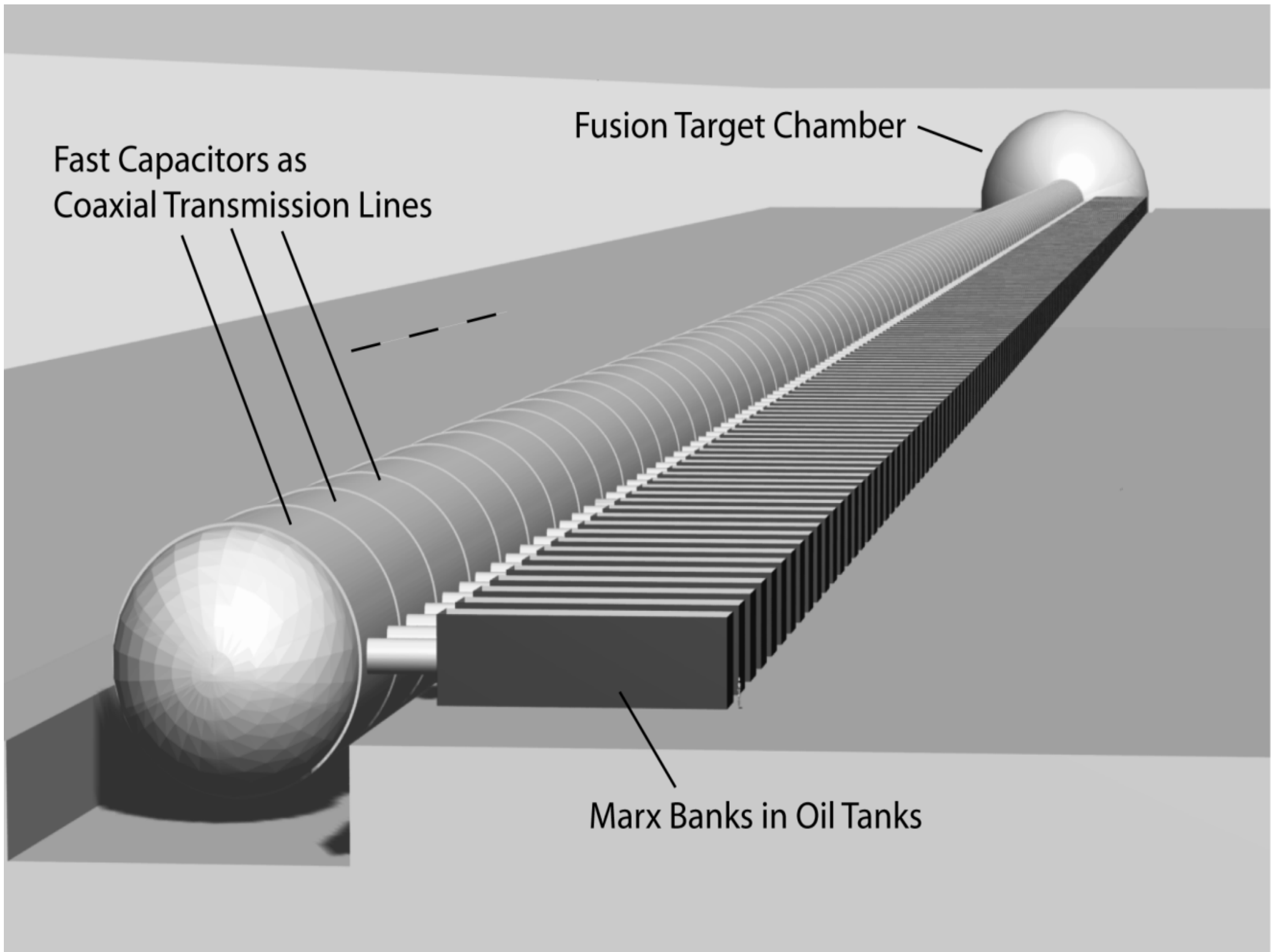

**Figure 9**: Artistic perception of a 1.5 km long Super Marx generator, composed of 100x 15 m long high voltage capacitors each designed as a magnetically insulated coaxial transmission line. The coaxial capacitors/transmission lines are placed inside a large vacuum vessel. Each capacitor/transmission line is charged by two conventional Marx generators up symmetrically to 10 MV (± 5 MV). After charge-up is completed, the Marx generators are electrically decoupled from the capacitors/transmission lines. The individual capacitors/transmission lines are subsequently connected in series via spark gap switches (i.e. the 'Super Marx' generator), producing a potential of 1 GV.

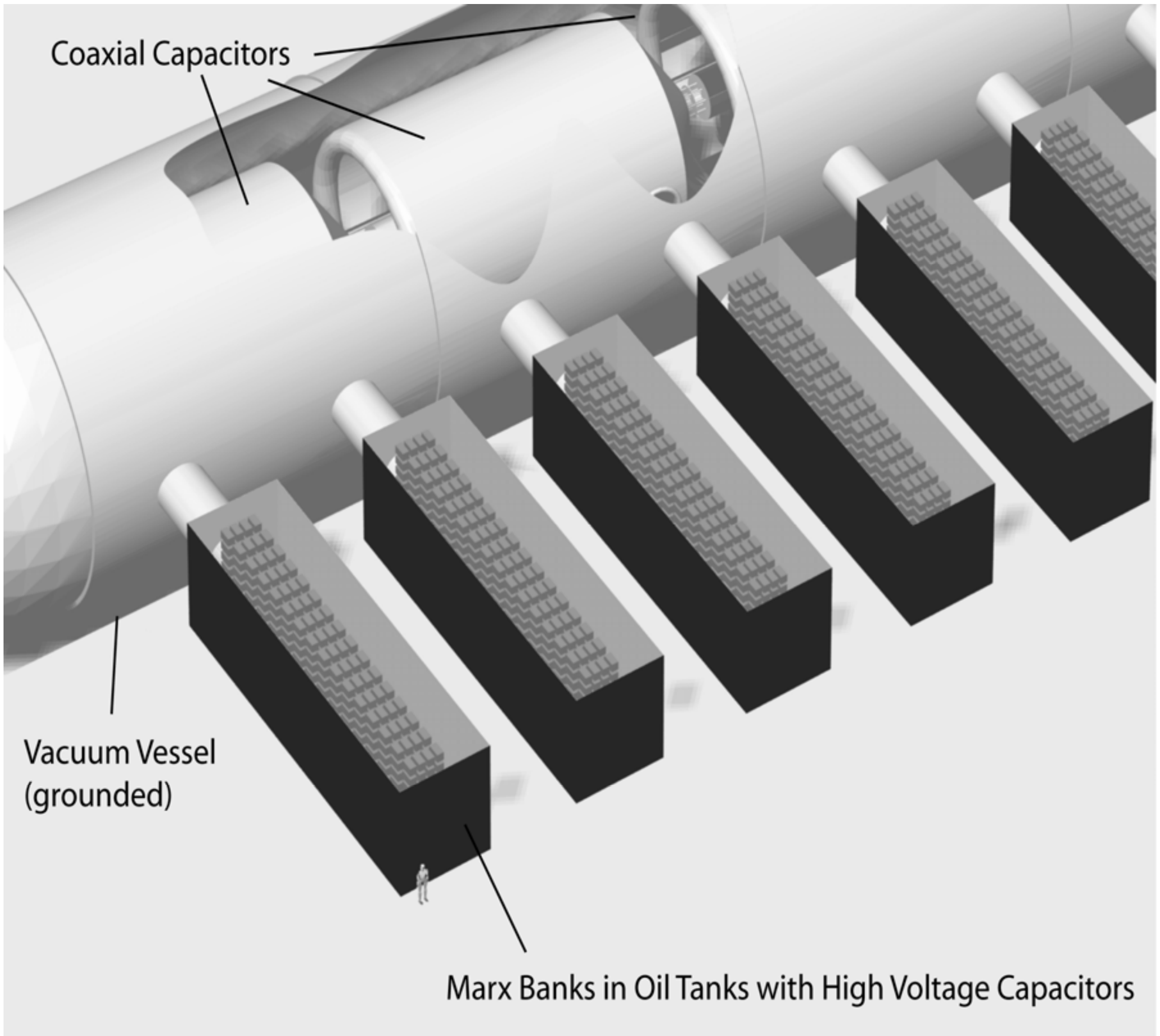

**Figure 10**: Detail view of a section of the Super Marx generator. Two conventional Marx banks charge up one coaxial capacitor/transmission line element to 10 MV.

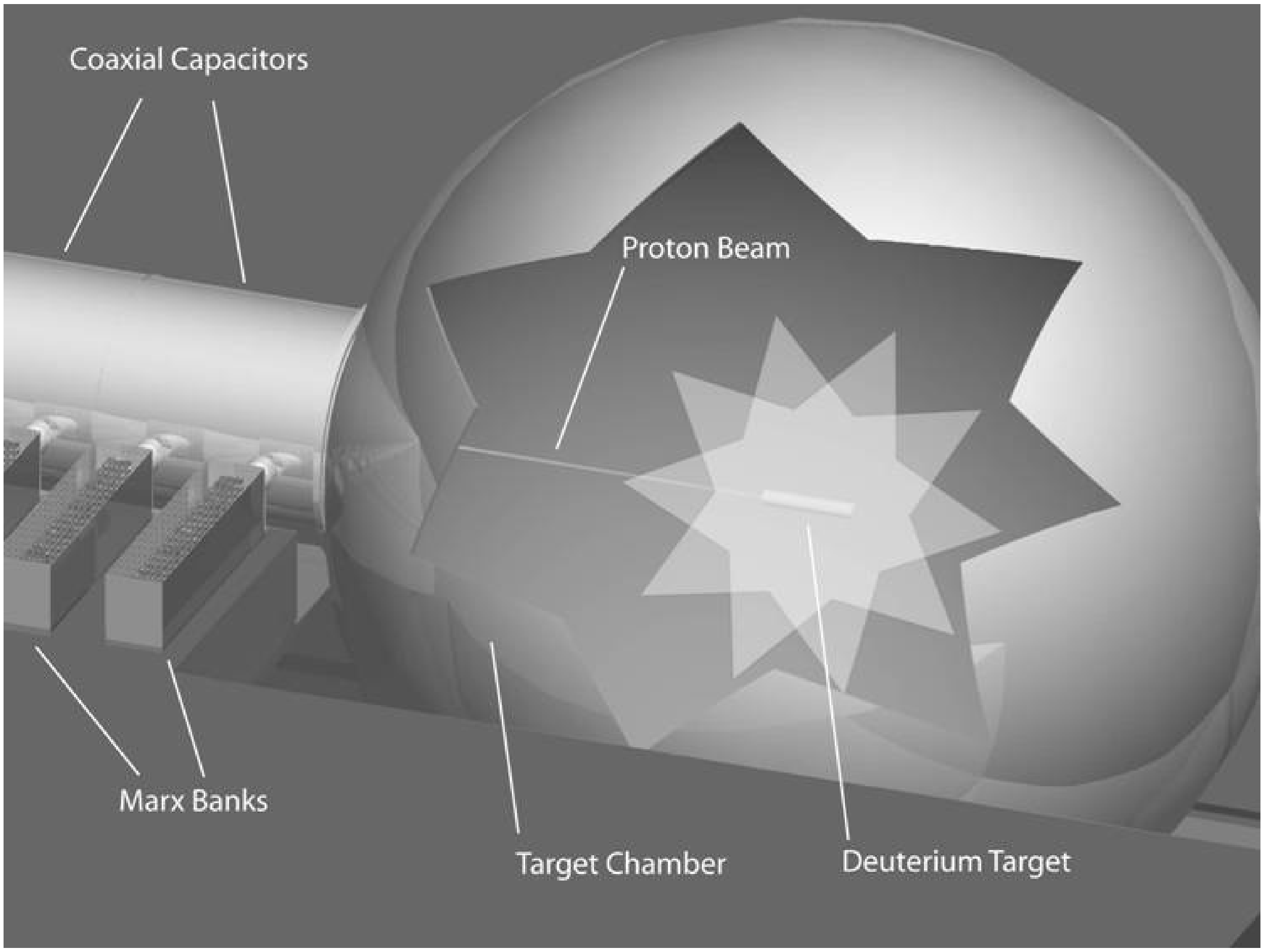

**Figure 11:** Injection of GeV – 10 MA proton beam, drawn from Super Marx generator made up of magnetically insulated coaxial capacitors into chamber with cylindrical deuterium target.

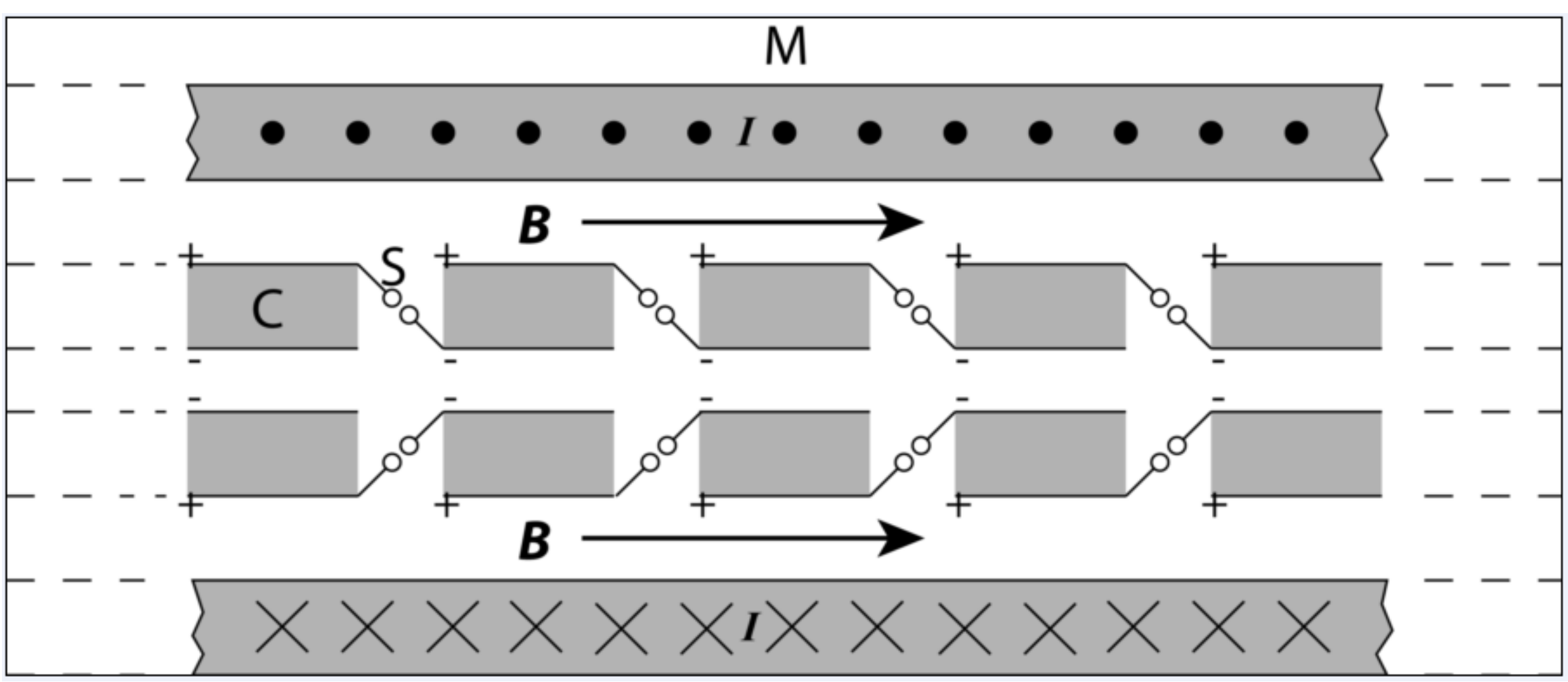

**Figure 12**: Showing a few elements of the Super Marx generator.

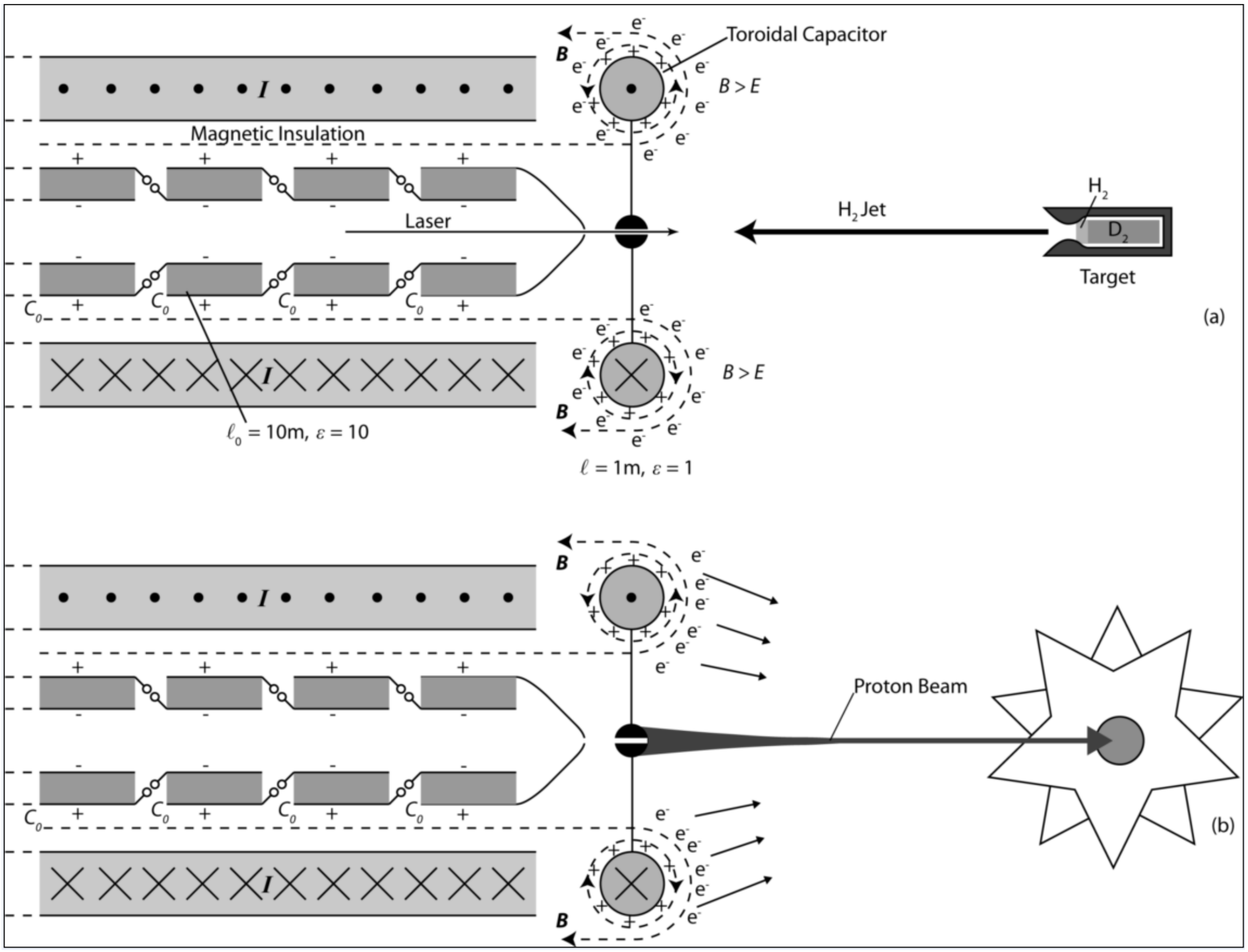

**Figure 13**: The superconducting toroidal capacitor (a) and its discharge onto the target (b)

which for the 100 capacitors of the Super Marx add up to $e \sim 10^{15}$ erg. About 10 times more energy can be stored, either if the radius of the capacitor is about 3 times larger, or with a larger dielectric constant, or with a combination of both. This means that for about 100 capacitors energy $10^{16}$ erg = 1 GJ can be stored in the mile-long Super Marx.

Another idea proposed by Fuelling [19], is to place the ordinary Marx generators of the 1st stage inside the coaxial capacitors of the Super Marx. The advantage of this configuration is that it does not require to disconnect the Marx generators from the capacitors of the Super Max prior to its firing. Because the charging and discharging of the Super Marx can there be done very fast, one can use compact water capacitors where $\varepsilon \cong 80$. And instead of magnetic insulation of the capacitors of the Super Marx against the outer wall one can perhaps use transformer oil for the insulation. Giving each inner segment of the Super Marx enough buoyancy, for example by adding air chambers, these segments can be suspended in the transformer oil. There the outer radius of the co-axial capacitors is much larger. This permits to store in the Super Marx gigajoule energies.

As shown in **Fig. 13**, the last capacitor of the Super Marx is a superconducting ring with a large toroidal currents. There, because of the large azimuthal magnetic field set up by the toroidal current magnetically insulates the ring against breakdown to the wall.

The load is the deuterium target shown in **Fig. 13**. It consists of a solid deuterium rod covered with a thin ablator placed inside a cylindrical hohlraum. To the left of the hohlraum and the deuterium rod is a mini-rocket chamber filled with solid hydrogen.

The method of discharging the energy from the Blumlein transmission line to the target then goes as follows (see **Fig. 13**):

1. A short laser pulse is projected into the mini-rocket chamber, through a hole of an electrode at the center of the ring.
2. By heating the hydrogen in the mini rocket chamber to a high temperature, a supersonic hydrogen jet is emitted through the Laval nozzle towards the electrode at the center of the ring, forming a bridge to the target.
3. A second laser pulse then traces out an ionization trail inside the hydrogen jet, facilitating an electric discharge to the target, with the space charge neutralizing plasma pinching the proton beam down to a small diameter.

The Super Marx generator therefore can accomplish on earth the same as the spacecraft acting as large capacitor can do in space.

The testing of an Argon ion laser driven by high explosives can, of course, be done on earth and the same applies to the conjectured super-explosives.

## 12. Conclusion

If large scale manned spaceflight has any future, a high specified impulse – high thrust propulsion system is needed. The only known propulsion concept with this property is the nuclear bomb propulsion concept. However, since large yield nuclear explosions are for obvious reason undesirable, the nuclear explosions should by comparison be small. But because of the "tyranny of the critical mass" (quote by F. Dyson), small fission bombs or fission triggered fusion bombs, become extravagant, with only a fraction of the nuclear material consumed[1].

In the original Orion bomb propulsion concept, the propulsive power was through the ablation of a pusher plate. There the energy is delivered to the pusher plate by the black body radiation of the exploding bomb. The propulsion by non-fission triggered fusion bombs, not only has the advantage that it is not subject to the "tyranny of the critical mass", but the propulsive power is there delivered by the kinetic energy of expanding hot plasma fire ball repelled from the spacecraft by a magnetic mirror. This is in particular true for a pure deuterium bomb, where in comparison to DT more energy is released into charged fusion products. In a DT bomb, 80% of the energy goes into neutrons.

While in a fission explosion most of the energy is lost into space by the undirected blackbody radiation, much more propulsive energy can be drawn from the plasma of a pure deuterium fusion bomb explosion, in conjunction with a magnetic mirror.

Manned space flight requires lifting large masses into earth orbit, where they are assembled into a large spacecraft. While this can be done with chemical rockets, it would be much more economical if it could be done with a chain of small nuclear explosions. Without radioactive fallout this can be done with a chain of laser ignited fusion bombs, with one laser for each bomb, where the lasers become part of the exhaust. Ignition cannot be done by infrared chemical or $CO_2$ lasers as it was suggested by the Los Alamos team [18], but rather by the kind of an ultraviolet laser driven by high explosives, as suggested by the author [9].

Looking ahead into the future with deuterium as the nuclear rocket fuel, widely available on most planets of the solar system and in the Oort cloud outside the solar system, this would make manned space flight to the Oort cloud possible, at a distance at about one tenth of one light year.

---

[1] The same is true for nuclear fission gas core rocket reactors, where much of the unburnt fission fuel is lost in the exhaust.

## Appendix: Conjectured Metastable Super-Explosives formed under high pressure for thermonuclear ignition

Under normal pressure the distance of separation between two atoms in condensed matter is typically of the order $10^{-8}$ cm, with the distance between molecules formed by the chemical binding of atoms of the same order of magnitude. As illustrated in a schematic way in **Fig. A1**, the electrons of the outer electron shells of two atoms undergoing a chemical binding, form a "bridge" between the reacting atoms. The formation of the bridge is accompanied in a lowering of the electric potential well for the outer shell electrons of the two reacting atoms, with the electrons feeling the attractive force of both atomic nuclei. Because of the lowering of the potential well, the electrons undergo under the emission of eV photons a transition into lower energy molecular orbits. At higher pressures, bridges between the next inner shells are formed, under the emission of soft X-rays.

Going to still higher pressures, a situation can arise as shown in **Fig. A2**, with the building of electron bridges between shells inside shells. There the explosive power would be even larger. Now consider the situation where the condensed state of many closely spaced atoms is put under high pressure making the distance of separation between the atoms much smaller, and where the electrons from the outer shells coalesce into one shell surrounding both nuclei, with the electrons from inner shells forming a bridge. Because there the change in the potential energy is much larger, the change in the electron energy levels is also much larger, and can be of the order of keV. There then a very powerful explosive is formed releasing its energy in a burst of keV X-rays. This powerful explosive is likely to be very unstable, but it can be produced by the sudden application of a high pressure in just the moment when it is needed. Because an intense burst of X-rays is needed for the ignition of a thermonuclear microexplosion, it could be used as an alternative to the argon ion laser for the ignition of a pure fusion bomb.The energy of an electron in the ground state of a nucleus with the charge Ze is

$$E_1 = -13.6Z^2 \text{ [eV]}. \tag{A.1}$$

With the inclusion of all the Z electrons surrounding the nucleus of charge Ze, the energy is

$$E_1^* \approx -13.62\, Z^{2.42} \text{ [eV]} \tag{A.2}$$

with the outer electrons less strongly bound to the nucleus.

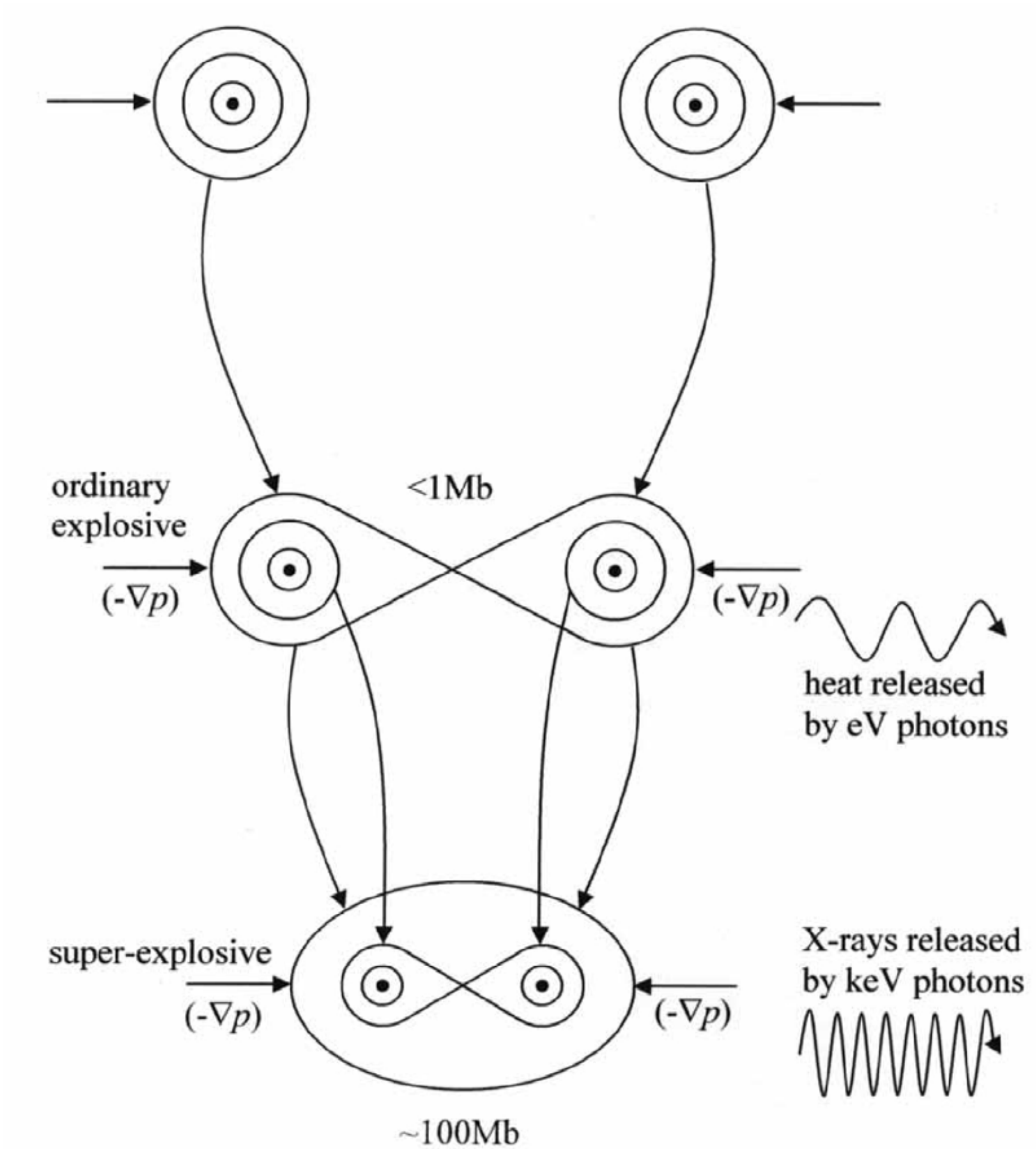

**Fig.A1** In an ordinary explosive the outer shell electrons of the reacting atoms form “eV” molecules accompanied by the release of heat through eV photons. In a superexplosive the outer shell electrons “melt” into a common outer shell with inner electron shells form “keV” molecules accompanied by the release of X-ray keV photons.

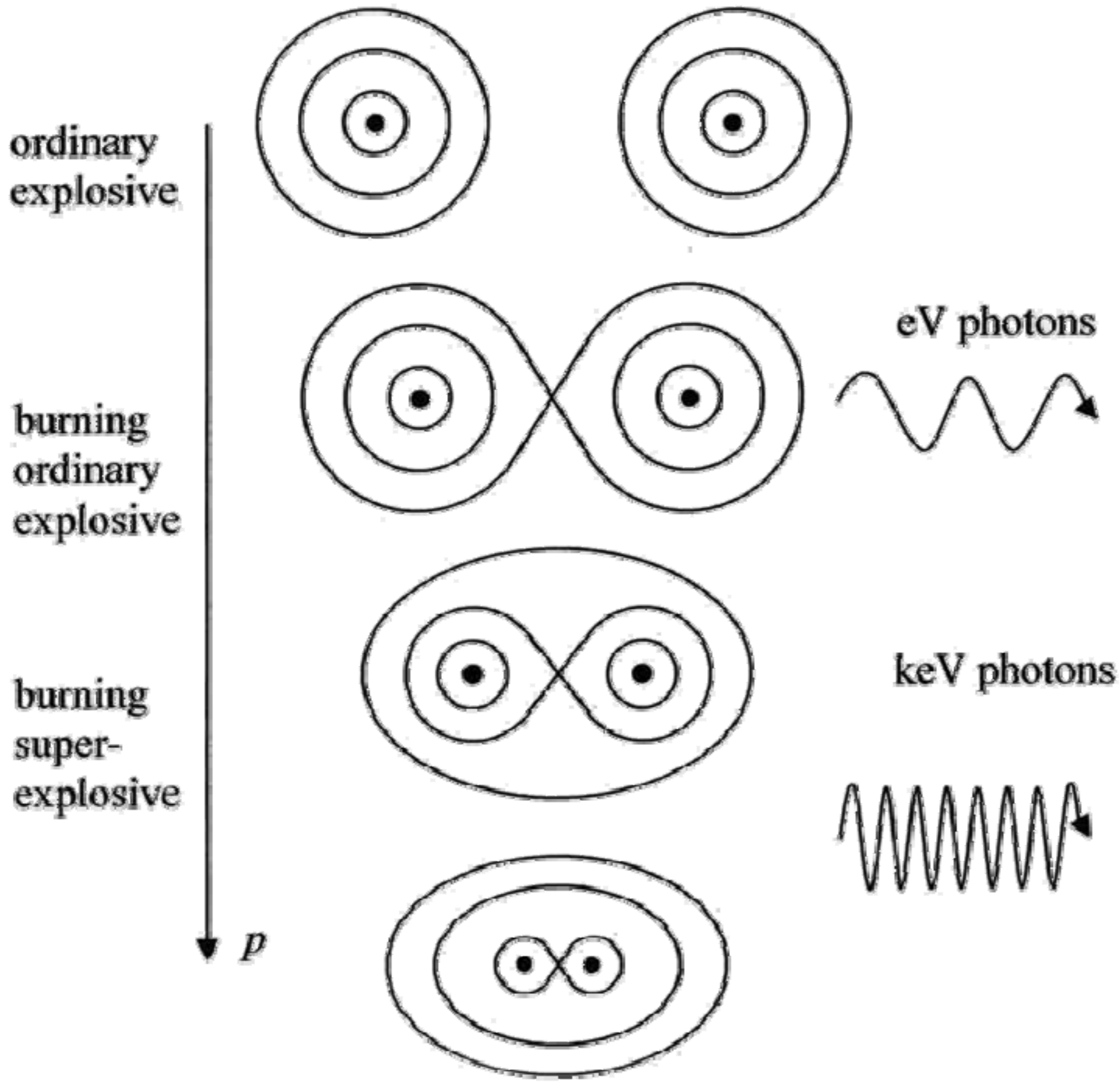

**Fig. A2**. With increasing pressure electron-bridges are formed between shells inside shells melting into common shells.

Now, assume that two nuclei are so strongly pushed together that they act like one nucleus with the charge 2Ze, onto the 2Z electrons surrounding the 2Ze charge. In this case, the energy for the innermost electron is

$$E_2 = -13.6(2Z)^2 \quad [\mathrm{eV}] \tag{A.3}$$

or if the outer electrons are taken into account,

$$E_2^* = -13.6\,(2Z)^{2.42} \quad [\mathrm{eV}] \tag{A.4}$$

For the difference one obtains

$$\delta E = E_1^* - E_2^* = 13.6Z^{2.42}\,(2^{2.42} - 1) \approx 58.5Z^{2.42}\ [\mathrm{eV}]. \tag{A.5}$$

Using the example Z = 10, which is a neon nucleus, one obtains $\delta E \approx 15$ keV. Of course,

it would require a very high pressure to push two neon atoms that close to each other, but this example makes it plausible that smaller pressures exerted on heavier nuclei with many more electrons may result in a substantial lowering of the potential well for their electrons.

A pressure of $p \approx 100$Mb = $10^{14}$ dyn/cm$^2$, can be reached with existing technology in sufficiently large volumes, with at least three possibilities:

1. Bombardment of a solid target with an intense relativistic electron- or ion beam.
2. Hypervelocity impact.
3. Bombardment of a solid target with beams or by hypervelocity impact, followed by a convergent shock wave.

**To 1:** This possibility was considered by Kidder [20] who computes a pressure of 50 Megabar (Mb), if an iron plate is bombarded with a 1 MJ – 10 MeV – $10^6$ A relativistic electron beam, focused down to an area of 0.1 cm$^2$. Accordingly, a 2 MJ beam would produce 100 Megabar. Instead of using an intense relativistic electron beam, one may use an intense ion beam. It can be produced by the same high voltage technique, replacing the electron beam diode by a magnetically insulated diode [21].

Using intense ion beams has the additional benefit that the stopping of the ions in a target is determined by a Bragg curve, generating the maximum pressure inside the target, not on its surface.

**To 2:** A projectile with the density $\rho \approx 20$ g/cm$^3$, accelerated to a velocity v = 30 km/s would, upon impact, produce a pressure of $p \approx 100$ Mb. The acceleration of the projectile to these velocities can be done by a magnetic traveling wave accelerator.

**To 3:** If, upon impact of either a particle beam or projectile, the pressure is less than 100 Mb, for example only of the order 10 Mb, but acting over a larger area, a tenfold increase in the pressure over a smaller area is possible by launching a convergent shock wave from the larger area on the surface of the target, onto a smaller area inside. According to Guderley , the rise in pressure in a convergent spherical shock wave goes as $r^{-0.9}$ , which means that 100 Megabar could be reached by a ten-fold reduction in the radius of the convergent shock wave.

While it is difficult to reach 30 km/s with a traveling magnetic wave accelerator, it is easy to reach a velocity of 10km/s with a two stage light gas gun.

We assume an equation of state of the form $p/p_0 = (n/n_0)^\gamma$. For a pressure of 100Mb=$10^{14}$ dyn/cm$^2$, we may set γ = 3 and $p_0$= $10^{11}$ dyn/cm$^2$, $p_0$ being the Fermi pressure of a solid at the atomic number density $n_0$, with $n$ being the atomic number density at the elevated pressure $p > p_0$. With $d = n^{-1/3}$, where $d$ is the lattice constant, one has

$$d/d_0 = (p/p_0)^{1/9} \tag{A.6}$$

For $p$ = $10^{14}$ dyn/cm$^2$. Such a lowering of the inneratomic distance is sufficient for the formation of molecular states.

Calculations done by Muller, Rafelski, and Greiner [22], show that for molecular states $_{35}$Br- $_{35}$Br, $_{53}$I- $_{79}$Au, and $_{92}$U- $_{92}$U, a two-fold lowering of the distance of separation leads to a lowering of the electron orbit energy eigenvalues by ~ 0.35 keV, 1.4 keV, respectively. At a pressure if 100 Mb = $10^{14}$ dyn/cm$^2$ where $d/d_0 \cong 1/2$, the result of these calculations can be summarized by (δE in keV)

$$log\, \delta E \cong 1.3 \times 10^{-2} Z - 1.4 \tag{A.7}$$

replacing eq. (A.5), where $Z$ is here the sum of the nuclear charge for both components of the molecule formed under the high pressure.

The effect the pressure has on the change in these quasi-molecular configurations is illustrated in **Fig. A3**, showing a $p – d$ (pressure-lattice distance) diagram. This diagram illustrates how the molecular state is reached during the compression along the adiabat $a$ at the distance $d = d_c$ where the pressure attains the critical value $p = p_c$. In passing over this pressure the electrons fall into the potential well of the two-center molecule, releasing their potential energy as a burst of X-rays. Following its decompression, the molecule disintegrates along the lower adiabat $b$.

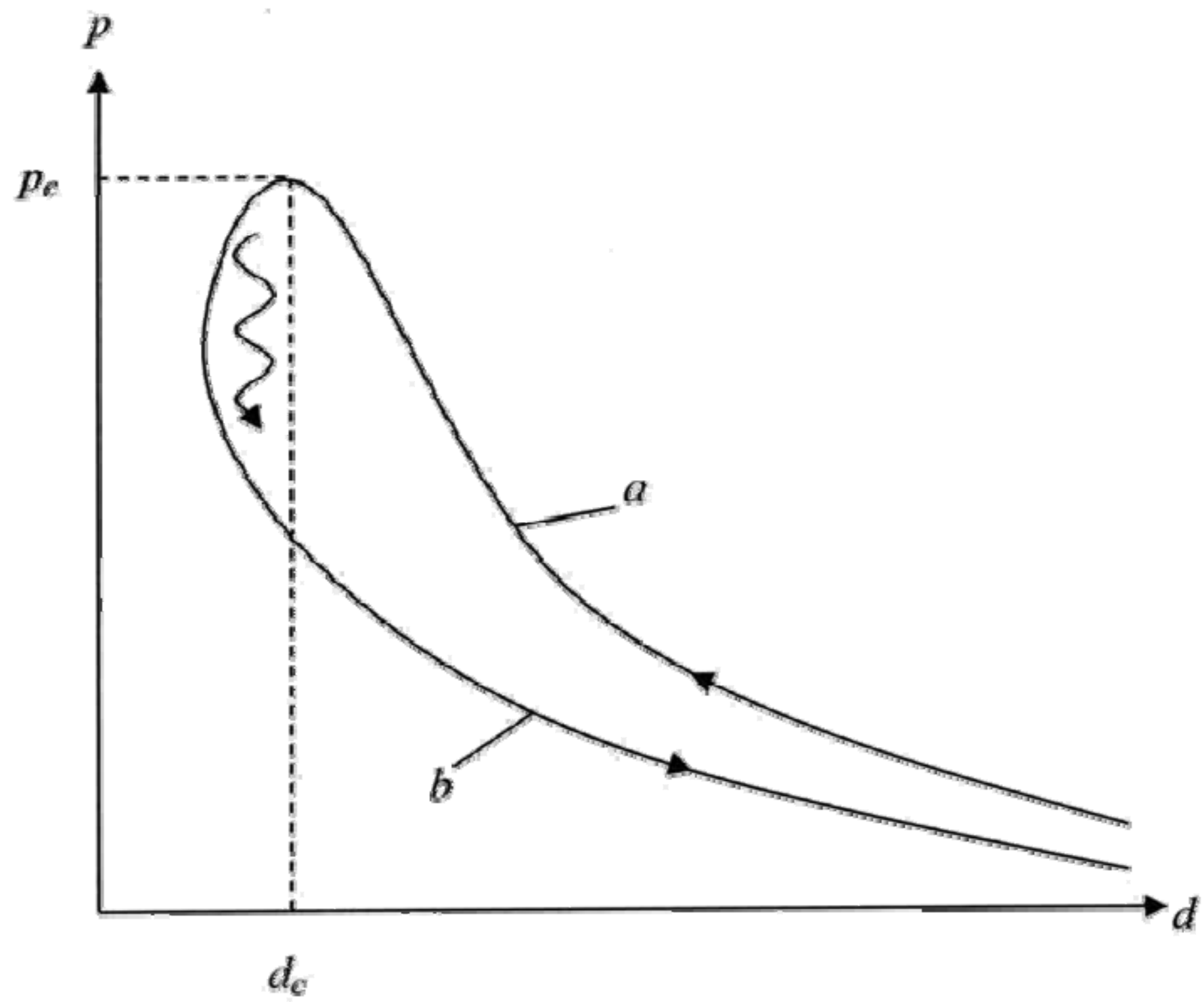

**Fig. A3**. *p-d,* pressure – inneratomic distance diagram for the upper atomic and lower molecular adiabat.

If the conjectured super-explosive consists of just one element, as is the case for the $_{35}$Br - $_{35}$Br reaction, or the $_{92}$U - $_{92}$U reaction, no special preparation for the super-explosive is needed. But as the example of Al–FeO thermite reaction shows, reactions with different atoms can release a much larger amount of energy compared to other chemical reactions. For the conjectured super-explosives this means as stated above that they have to be prepared as homogeneous mixtures of nano-particle powders, bringing the reacting atoms come as close together as possible.

For the ignition of a thermonuclear reaction one may consider the following scenario illustrated in **Fig. A4**. A convergent shock wave launched at the radius $R = R_0$ into a spherical shell of outer and inner radius $R_0$ , $R_1$ , reaches near the radius R = $R_1$ at a pressure of 100 Mb. After the inward moving convergent shock wave has reached the radius $R = R_1$, an outward moving rarefaction wave is launched from the same radius $R = R_1$ , from which an intense burst of X-rays is emitted. One can then place a thermonuclear DT target inside the cavity of the radius $R = R_1$ , with the target bombarded, imploded, and ignited by the X-ray pulse. The ignited DT can there serve as a “hot spot” for the ignition of deuterium.

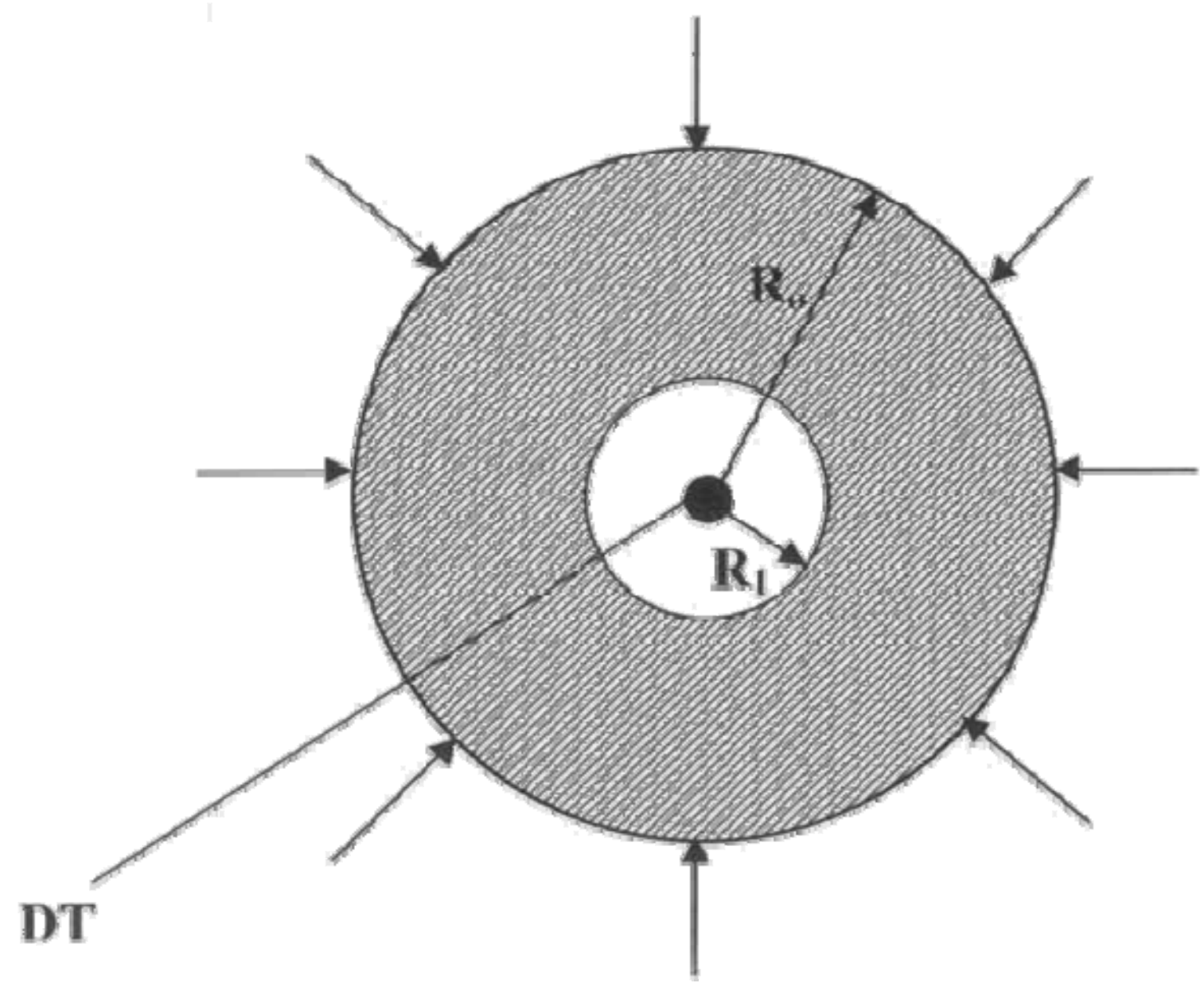

**Fig. A4**. Inertial confinement fast ignition configuration

## Acknowledgement

In writing this report I acknowledge the encouragement of the following individuals listed in alphabetical order: Robert Bigelow, Stephen Fuelling, Kelvin Long, Harold Puthoff and Harry Ruppe.